\documentclass[usenatbib]{mn2e} 

\usepackage{mathtools}
\usepackage{graphicx}
\usepackage{epstopdf}
\usepackage{natbib}
\usepackage{verbatim}

 \usepackage{times}
\renewcommand{\thefootnote}{\fnsymbol{footnote}}

\newcommand{\apj}{ApJ} \newcommand{\apjl}{ApJL}
\newcommand{\aap}{A\&A} 
\newcommand{\aj}{AJ} \newcommand{\apjs}{ApJS} 
\newcommand{\mnras}{MNRAS} \newcommand{\nat}{Nature}
 
\newcommand{\procspie}{Proc. SPIE }

\def\lsim{\mathrel{\rlap{\lower4pt\hbox{\hskip1pt$\sim$}}
    \raise1pt\hbox{$<$}}}                % less than or approx. symbol
\def\gsim{\mathrel{\rlap{\lower4pt\hbox{\hskip1pt$\sim$}}
    \raise1pt\hbox{$>$}}}                % greaterthanorapprox. symbol

\begin{document}

\title[Spectroscopically-confirmed gravitationally lensed galaxies in SDSS]
{The CASSOWARY spectroscopy survey: A new sample of gravitationally lensed galaxies in SDSS}

\author[Stark et al.]  {Daniel P. Stark$^{1}$\footnotemark[2],
Matthew Auger$^{2}$, 
Vasily Belokurov$^{2}$, 
Tucker Jones$^{3}$, 
Brant Robertson$^{1}$,
\newauthor
Richard S. Ellis$^{4}$, 
David J. Sand$^{5}$,
Alexei Moiseev$^{6}$,
Will Eagle$^{2}$,
Thomas Myers$^{2}$
  \\
$^{1}$ Steward Observatory, University of Arizona, 933 N Cherry Ave, Tucson, AZ, USA \\  
$^{2}$ Institute of Astronomy, University of Cambridge, Madingley Road, Cambridge CB3 0HA, United Kingdom\\
$^{3}$ Department of Physics, University of California, Santa Barbara  CA 93106, USA\\
$^{4}$ Department of Astronomy, California Institute of Technology, Pasadena, CA 91125 USA \\
$^{5}$ Department of Physics, Texas Tech University, Box 41051, Lubbock, TX 79409 USA \\
$^{6}$ Special Astrophysical Observatory, Nizhniy Arkhyz, Karachaevo-Cherkessiya, Russia \\
}
\date{Accepted ... ;  Received ... ; in original form ...}

\pagerange{\pageref{firstpage}--\pageref{lastpage}} \pubyear{2013}

\hsize=6truein
\maketitle

\label{firstpage}
\begin{abstract}
Bright gravitationally lensed galaxies provide our most detailed view of galaxies at high redshift. 
The very brightest ($r<21$) systems enable high spatial and spectral resolution measurements, offering 
unique constraints on the outflow energetics, metallicity gradients, and stellar populations 
in high redshift galaxies.  Yet as a result of the small number of ultra-bright $z\simeq 2$ lensed systems 
with confirmed redshifts, most detailed spectroscopic studies have been limited in their scope.  With the goal of 
increasing the number of bright lensed galaxies available for detailed follow-up, we have 
undertaken a spectroscopic campaign targeting wide separation ($\gsim $ 3\arcsec) galaxy-galaxy 
lens candidates within the Sloan Digital Sky Survey (SDSS).   Building on the earlier efforts of our 
CASSOWARY survey, we target a large sample of candidate galaxy-galaxy lens systems in SDSS using 
a well-established search algorithm which identifies blue arc-like structures situated around 
luminous red galaxies.    In this paper, we present a new redshift catalog containing 25 lensed sources in 
SDSS confirmed through spectroscopic follow-up of candidate galaxy-galaxy lens systems.   Included in 
this new sample are two of the brightest galaxies ($r=19.6$ and 19.7) galaxies 
known at $z\simeq 2$, a  low metallicity (12 + log (O/H) $\simeq 8.0$) extreme nebular line 
emitting galaxy at $z=1.43$, and numerous 
systems for which detailed follow-up will be possible.  The 
source redshifts span $0.9<z<2.5$ (median redshift of 1.9), 
and their optical magnitudes are in the range $19.6\lsim r\lsim 22.3$.    We present a brief 
source-by-source discussion of the 
spectroscopic properties extracted from our  confirmatory spectra and discuss some initial science results.   Preliminary lens modelling reveals 
average source magnifications of 5-10$\times$.    With more than 50 gravitationally-lensed $z\gsim 1$ galaxies now 
confirmed within SDSS, it will soon be possible for the first time to develop generalised conclusions from 
detailed spectroscopic studies of the brightest lensed systems at high redshift.

\end{abstract}


\renewcommand{\thefootnote}{\fnsymbol{footnote}}
\footnotetext[1]{E-mail: dpstark@email.arizona.edu}
\footnotetext[2]{Hubble Fellow.}

\section{Introduction}

Recent years have seen significant progress in the characterisation of star formation and 
feedback in high redshift galaxies.   While much of these strides have stemmed from spectroscopic 
study of large  samples of $z\simeq 2-4$ galaxies  (e.g., Shapley 2011 for a recent review), strong 
gravitational lensing is now playing an increasingly important role.  As typical galaxies are often 
too faint and too small to study individually in great detail, the magnification provided by strong 
gravitational lensing (often 10-30$\times$) provides one of our only means of dissecting high 
redshift galaxies in great detail with current facilities.

Study of strongly-lensed gravitationally lensed galaxies helps high-redshift studies in two ways. 
First the flux amplification enables spectra of higher resolution and increased signal to noise, yielding 
unique constraints on the absorbing gas 
(Pettini et al. 2002, Quider et al. 2009, 2010, Dessauge-Zavadsky et al. 2010) and detailed 
emission line studies of the ionised gas (e.g., Fosbury et al. 2003, Yuan et al. 2009, Hainline et al. 
2009, Bian et al. 2010, Rigby et al. 2011, Richard et al. 2011, Wuyts et al. 2012, Brammer et al. 2012).  
Second, the increase in the apparent size of the lensed sources enables characterisation of the 
internal kinematics and metallicity gradients on scales as small as 100 parsec (e.g., Stark et 
al. 2008, Swinbank et al. 2009, Jones et al. 2010a, 2010b, 2012b).

A tremendous amount of physical information becomes accessible through analysis of the 
optical spectra of the  brightest lensed galaxies.  The wide 
array of interstellar absorption features that are detectable in the rest-frame UV of bright lensed galaxies 
enable abundances to be measured for numerous 
chemical elements.   In a detailed study of MS 1512-cB58, one of the highest surface 
brightness lensed galaxies, Pettini et al. (2002) demonstrated that while the ISM is already heavily enriched 
with elements released in Type II supernovae, those elements (N and Fe-peak) produced on longer 
timescales (via intermediate mass stars and Type Ia supernovae) are significantly underabundant.    
These results provide a unique measure of the past star formation history, indicative of a 
picture whereby the bulk of metal enrichment has occurred in the past $\simeq 300$ Myr.   
Whether these abundance patterns vary with redshift, galaxy mass, or star formation rate 
remains unclear owing to the small sample of galaxies bright enough for such detailed analysis.   

The absorption line measurements of ultra-bright lensed galaxies additionally provide our most 
complete picture of the kinematics and geometry of the interstellar gas.    In the handful of lensed systems with 
detailed measurements, absorption is seen over a velocity range of 1000 km s$^{-1}$ with respect to the 
stars.    As expected, the bulk of the interstellar material in these systems is outflowing from the 
stars with typical speeds of 100-300 km s$^{-1}$.    In the case of MS 1512-cB58, the rate at which mass 
is ejected in the outflow is shown to be comparable if not greater than rate at which gas is converted to stars 
(Pettini et al. 2000, 2002), consistent with recent inferences from the detection of a broadened component 
of H$\alpha$ emission lines in unlensed $z\simeq 2$ galaxies from the SINS survey (Genzel et al. 2011, 
Newman et al. 2012).    Further constraints on the outflow mass loading factor are of crucial importance 
to an understanding of how outflows regulate star formation and the chemical evolution of high redshift galaxies. 

With the development of adaptive-optics assisted infrared integral field spectrographs 
on 8-10m telescopes, similarly unique insights are being achieved through 
measurement of the distribution of resolved properties within lensed galaxies at $z\simeq 2-3$.  
Most recently,  resolved metallicity gradient measurements have been presented for five 
bright lensed galaxies (Jones et al. 2010b, 2012b; Yuan et al. 2011) with source plane resolution as 
fine as 300 pc.   The current data point toward 2-3$\times$ steeper metallicity gradients at 
higher redshift, as expected in realistic models of inside-out growth (see Jones et al. 2012b); 
yet with the small existing samples of lensed galaxies with resolved metallicity measurements, significant uncertainties in the 
average gradient evolution remain.

Clearly the potential impact of lensing studies has been stunted by the small number of the 
brightest systems for which detailed analysis is possible.   With only a handful of 
galaxies having absorption line measurements with adequate S/N and resolution, it is difficult to assess population averages 
of the mass loading factor and ISM chemical abundance patterns.  Near-IR studies are more severely 
compromised, as many of the brightest systems happen to be located 
at redshifts which place their strong rest-optical emission lines behind sky lines.  
Larger samples of bright $z\simeq 2-3$ gravitationally-lensed galaxies are required to build upon the 
emerging framework.    Fortunately,  through 
a variety of imaging surveys ranging from the Sloan Digital Sky Survey (SDSS) to  
HST surveys of galaxy clusters (e.g., Ebeling et al. 2010), the number of bright 
lensed galaxies has rapidly increased in recent years.   

Our group is one of several (see also e.g., Hennawi et al. 2008; Diehl et al. 2009; Kiester et al. 2010; Kubo et al.  2011;  Bayliss 
et al. 2011a, 2011b; Bayliss 2012) that has actively exploited SDSS imaging for the purposes of 
locating galaxy galaxy lenses.   As part of the Cambridge And Sloan Survey 
Of Wide ARcs in theskY; CASSOWARY), we have developed a search algorithm 
which targets multiple blue companions (or arcs) separated by $\gsim 3$ arcseconds from 
luminous red galaxies (Belokurov et al. 2009).   The initial search (Belokurov et al. 2007) yielded 
discovery of SDSS J1148+1930 (the "Cosmic Horseshoe"), a highly-magnified $z=2.38$ 
galaxy bright enough (r=19.5) for high resolution optical spectroscopy (Quider et al. 2009) 
and emission line mapping (Jones et al. 2012b), while subsequent exploration 
has yielded further candidates (e.g., Belokurov et al. 2009; Pettini et al. 2010; Christensen et al. 2010; 
Brewer et al. 2011).   Typically the CASSOWARY systems correspond to lensing of $z \simeq 1-3$
star-forming galaxies by an early type galaxy and fainter group 
companions at $z \simeq 0.2$ -  0.7.  

Yet in spite of this progress, there still remains nearly 50 bright candidate lensed systems 
in our photometric catalog (as well as catalogs of other teams, see Wen et al. 2011) without confirmed redshifts. 
Among those lacking confirmation are potentially some of the brightest gravitationally-lensed galaxies 
on the sky.    To help build a more complete redshift database of the optically-brightest lensed galaxies 
in the SDSS footprint,  we have undertaken a spectroscopic campaign (primarily with MMT, 
but also using Keck, Magellan, and LBT)  to obtain redshifts for bright lensed candidates in our CASSOWARY 
catalog.  Through these efforts, we present spectroscopic redshifts for 25 lensed CASSOWARY 
sources at $0.9<z<2.8$ detected within SDSS.     As much of our focus is on the spatially-resolved and 
high spectral resolution study of high redshift galaxies (e.g., Stark et al. 2008; Quider et al. 2009, 2010; 
Jones et al. 2010a, 2010b, 2012b), we put significant effort into confirming those targets which are 
optimal for this study.   Follow-up spectroscopy of the new CASSOWARY sample is under way and will be 
described in future papers.

The plan of the paper is as follows.  In \S2, we review the lens search algorithm and describe 
the observations undertaken for this program.     In \S3, we present the new redshift sample, providing a brief 
discussion of the spectroscopic properties of each new system.   In \S4, we describe  
preliminary lens models for select systems and comment on the typical star formation 
rates and stellar masses of the new CASSOWARY sample.   In \S5, we  
provide a more extended discussion of two lensed systems and describe some of the opportunities 
that will be made possible with the new sample.

Throughout the paper, we adopt a $\Lambda$-dominated, flat universe
with $\Omega_{\Lambda}=0.7$, $\Omega_{M}=0.3$ and
$\rm{H_{0}}=70\,\rm{h_{70}}~{\rm km\,s}^{-1}\,{\rm Mpc}^{-1}$. All
magnitudes in this paper are quoted in the AB system (Oke et al. 1983).

\section{Lens Identification and Observations}
 
We began a spectroscopic campaign of CASSOWARY candidate lensed galaxies in 
September of 2011.   Below we review our SDSS lens search algorithm and then provide 
details of the instrumentation used on each of these facilities.   
An overview of the observations is provided in Table 1.

\subsection{Lens search algorithm}

The CASSOWARY survey (Belokurov et al. 2007, 2009) is designed to  find  
gravitational lenses with angular Einstein diameters in excess of the size of the SDSS fibre ( $\gsim 3$\arcsec).   The motivation 
of the survey and the search procedure is described in detail in Belokurov et al. (2009).   Here we provide a short summary.  

Our lens search algorithm relies primarily on the SDSS photometric catalogues, rather than the actual images. The basic 
idea is to find massive elliptical galaxies surrounded by smaller objects with the spectral energy distributions (SEDs) 
suggestive of star-forming galaxies at $1 < z < 3$. The algorithm first identifies "seed" lenses by selecting 
massive Luminous Red Galaxies (LRGs) which we inspect for evidence of lensing.    The LRG selection uses 
a modified version of the criteria presented in Eistenstein et al. (2001), allowing for slightly lower surface brightness, fainter apparent magnitude and slightly bluer colours.   The SDSS photometric quality flags are used to cull the artefacts produced 
by severe blending and saturated stars. 

In the second step, using the SDSS Neighbours list for each of our "seed lenses", we identified 
systems with more than 1 blue neighbour within 10-20\arcsec of the lens candidate. The exact criteria of how blue the arc candidates ought to be 
were tuned for different iterations of the search, but typically we require at least $(g-r)_S < 1$ or $(g-r)_S < (g-r)_L$ where "S" refers to the 
blue lensed source candidate and "L" corresponds to the red lens candidate.  

The final step is to filter out false positives. There are several astrophysical phenomena other than 
strong gravitational lensing that could pass the first two
filters. These include nearby spiral galaxies with prominent bulges, polar rings, and various 
configurations of minor mergers. We take 
advantage of the existing SDSS spectroscopy to filter out the majority of the nearby spirals. 
The polar rings and the mergers are picked out
 by visual inspection of the SDSS image cutouts. The final result of this procedure is a sample of 
 $\gsim 100$ CASSOWARY gravitational lens candidates.  We use this as the input catalog for our 
 spectroscopic search described below.  

\subsection{MMT}

Spectroscopic observations with MMT were conducted in longslit mode primarily with the Blue Channel Spectrograph.   
We used the 300 line grating of the Blue Channel Spectrograph, providing close to 5300~\AA\ of spectral coverage.   
We typically used a central wavelength of  5500~\AA\ with a UV-36 order blocking filter.   We adopted a 1 arcsec slit width, providing a spectral resolution of 6.5~\AA.   Observations with the Blue Channel Spectrograph are optimal for confirming $z\gsim 2$ galaxies via detection of UV absorption lines.   

In the redshift range $1.0<z<1.5$, it is far easier to confirm star-forming lensed galaxies through 
detection of [OII] emission in the red.   We have thus 
also conducted observations with the MMT Red Channel Spectrograph.   We used 
the 270 line/mm grating with a central wavelength in the range $\sim 7500-7900$~\AA.     As with our 
blue channel observations, we use a slit width of 1 arcsec, providing a 
spectral resolution of $\sim 11$ ~\AA.   The spectral coverage with this setting is 
3705~\AA, allowing detection of emission lines out to $\simeq 9300-9700$~\AA, depending 
on the central wavelength.   During our red channel runs, we occasionally targeted candidate $z\simeq 2$ 
lensed galaxies.   In order to search for absorption from Si II $\lambda$1526, CIV $\lambda$1549, and Al II 
$\lambda$1671 in these $z\simeq 2$ sources, we 
used the 300 line/mm grating (with central wavelength of 5500~\AA), providing useful 
coverage in the range 
4230-7000~\AA.

Exposure times varied from candidate to candidate depending on source brightness 
and conditions.   We generally observed an individual galaxy for between 30 and 60 minutes.  
Specifics for each source are provided in Table 1.   The conditions were typically  
clear but not always photometric.   Seeing ranged from 0\farcs5-1\farcs5, but for most 
successful observations, the average seeing was $\lsim 1$\arcsec.
 
Reduction of blue and red channel longslit spectra were conducted using standard routines.   
Wavelength calibration was performed with HeAr/Th arcs.  Each spectrum was flux calibrated 
using spectral observations of standard stars.  
 
\subsection{Magellan}
  
We obtained near-infrared spectra for several of the CASSOWARY systems using the 
Folded-port InfraRed Echellette (FIRE; Simcoe et al. 2010) on the 6.5 meter Magellan Baade 
Telescope.    We used FIRE in echelle mode, delivering continuous spectral 
coverage between the z-band and K-band.    The slit length of FIRE is 7 arcseconds.   
We adopted a slit width of 0.75 arc seconds, providing a resolving power of 
$\rm{R=4800}$.    The FIRE data were reduced using the custom FIREHOSE pipeline.    
Wavelength calibration was performed using OH sky lines.    
We have thus far secured emission lines for four of the galaxies in the sample presented in this paper.  
Details are listed in Table 1 and 2.   
   
\subsection{Keck}

We used the DEIMOS spectrograph (Faber et al. 2003) on the Keck II telescope to confirm 
redshifts of several of our CASSOWARY candidates.   These observations were typically 
conducted during runs focused on deep observations of $z\simeq 6$ galaxies.  When conditions were not sufficient for such 
deep observations, we observed bright CASSOWARY targets.    We used the 1200 line/mm
grating which has a spectral resolution of $\simeq 1.1$~\AA.   These observations were 
focused in the red, providing spectral coverage of 2630~\AA\ in the wavelength range 
of 7000 to 10000 ~\AA.   Reduction was conducted using the standard spec2d DEIMOS pipeline.   
In total, we confirmed the redshifts of four lensed galaxies with 
DEIMOS.   Further details are in Table 1.

\begin{table*}
\begin{tabular}{lclcllccccll}
\hline  Lens Name & CSWA-ID & Observatory & Instrument  & Configuration & Dates  &$\rm{t_{exp}}$ (ksec)  &  PA (deg) \\  \hline 

SDSS J0058$-$0722 & 102 & MMT  & BCS  & 300 line/mm & 2011 Sep 29 &  3.6 & 340    \\  
\ldots 	& \ldots & MMT  & RCS  & 270 line/mm & 2013 Jan 21 &  1.8 &  85 \\  
\ldots 	& \ldots & MMT  & RCS  & 300 line/mm & 2013 Jan 22 &  4.8 &  -41    \\  
SDSS J0105+0145 &165 & MMT  & BCS  & 300 line/mm & 2011 Sep 30 & 2.7 &  2.5   \\  
... & ... & MMT  & BCS  & 300 line/mm & 2012 Dec 12 &  3.6 &  5.0   \\  
SDSS J0143+1607 & 116 & MMT  & BCS  & 300 line/mm & 2011 Sep 29 &  2.7 & 105    \\  
$\ldots$ & $\ldots$ & MMT  & RCS  & 270 line/mm &  2012 Nov 23 & 1.2   & 5     \\  
SDSS J0145$-$0455 	& 103 & MMT  & BCS  & 300 line/mm & 2011 Sep 29 &  3.6 & 340    \\  
SDSS J0232$-$0323 &164 & MMT  & BCS  & 300 line/mm & 2011 Sep 29 &  3.6 & 330    \\  
$\ldots$ & ...    & Magellan  & FIRE  & echelle & 2012 Feb 15 &  2.7 & 58    \\  
SDSS J0800+0812 &11 & MMT  & BCS  & 300 line/mm & 2012 Mar 24 &  1.8 & 60    \\  
\ldots & \ldots & MMT  & RCS  & 270 line/mm & 2012 May 01  & 2.7  & 60    \\  
SDSS J0807+4410 &139 & MMT  & BCS  & 300 line/mm & 2012 Mar 23 &  3.6 & 80    \\  
SDSS J0846+0446 & 141 & MMT  & BCS  & 300 line/mm & 2011 Nov 02 &  0.6 & 10    \\  
$\ldots$ &...     & Keck  & DEIMOS  & 1200 line/mm & 2011 Oct 26 &  2.7 & 10    \\  
$\ldots$ &...     & Magellan  & FIRE  & echelle & 2012 Feb 15 &  1.2 & 280    \\  
SDSS J0854+1008 &142 & Keck & DEIMOS  & 1200 line/mm & 2011 Oct 26 &  1.2 & 30    \\  
$\ldots$ &...    & MMT & BCS  & 300 line/mm & 2011 Nov 02 &  2.7 & 30   \\  
SDSS J0921+1810  & 31 & Keck & DEIMOS  & 1200 line/mm & 2011 Oct 26 &  2.7 & 120    \\  
\ldots  & \ldots & MMT & RCS  & 270 line/mm & 2012 Oct 21 &  2.7 & 120    \\  
SDSS J1002+6020 &117 & MMT  & BCS  & 300 line/mm & 2011 Nov 02 &  2.7 & 90   \\  
$\ldots$  & $\ldots$ &  Keck & DEIMOS & 1200 line/mm & 2011 Oct 26 &  1.2 & 90   \\  
$\ldots$  & $\ldots$ &  MMT & RCS & 270 line/mm & 2013 Jan 22 & 2.7 & -94 \\  
SDSS J1009+1937 &15 & MMT  & BCS  & 300 line/mm & 2012 Mar 24 &  2.7 & 55    \\  
SDSS J1110+6459 &104 & MMT  & BCS  & 300 line/mm & 2011 Nov 02 &  2.7 & 104    \\  
\ldots & \ldots & MMT & BCS  & 300 line/mm & 2012 Mar 24 &  2.4 & 104  \\
SDSS J1111+5308 &16 & MMT  & BCS  & 300 line/mm & 2012 Mar 24 &  2.7 & 267    \\  
SDSS J1115+1645 & 105 &  Magellan  & FIRE  & echelle & 2012 Feb 15 &  2.7 & 85.2    \\  
\ldots & \ldots &  MMT  & RCS   &   270 line/mm & 2012 Nov 22 &  0.6 &  -130   \\  
SDSS J1138+2754 &17 & MMT  & BCS  & 300 line/mm & 2012 Mar 24 &  2.4 & 85.2    \\  
\ldots & \ldots & MMT & RCS & 270 line/mm & 2012 May 01 &  2.7 & 85.2    \\  
SDSS J1147+3331 &107 &  MMT  & BCS  & 300 line/mm & 2012 Mar 23 &  3.6 & 280    \\  
\ldots & \ldots &  MMT  & RCS  & 270 line/mm & 2012 May 01 &  2.7 & 280    \\  
SDSS J1156+1911 &108 &  MMT  & BCS  & 300 line/mm & 2012 Mar 23 &  3.6 & 280    \\  
\ldots & \ldots &  MMT  & RCS  & 270 line/mm & 2013 Jan 21  &  3.6 & -20   \\  
SDSS J1237+5533 &13 & MMT  & BCS  & 300 line/mm & 2012 Mar 24 &  1.8 & 320   \\  
\ldots  & \ldots & MMT  & RCS  & 300 line/mm & 2012 July 06 &  1.8 & 320   \\  
\ldots & \ldots &  MMT  & RCS  & 300 line/mm & 2013 Jan 21 &  3.6 & 280    \\  
SDSS J1439+3250 &109 &  MMT  & BCS  & 300 line/mm & 2012 Mar 23 &  2.7  &  100    \\  
\ldots &\ldots &  MMT  & RCS  & 270 line/mm & 2013 Jan 22  &  3.6  & -10  \\  
\ldots &\ldots &  MMT  & RCS  & 300 line/mm & 2013 Jan 22    &   2.4  & -10 \\  
SDSS J1958+5950 &128 & MMT  & BCS  & 300 line/mm & 2011 Sep 30 &  2.7 & 56   \\  
\ldots & \ldots & MMT  & RCS  & 1200 line/mm & 2012 Jul 06 &  18 & 56   \\  
\ldots & \ldots & LBT  &  LUCI & 200\_H+K (HK) & 2012 Nov 06 &  3.6  &  -130   \\  
\ldots & \ldots & LBT  &  LUCI & 200\_H+K (zJ) & 2012 Nov 06 &  2.4  &  -130   \\  
SDSS J2158$+$0257  &163 & MMT  & BCS  & 300 line/mm & 2011 Sep 30 &  1.8 & 20    \\  
\ldots                        & & Magellan  & FIRE  & echelle & 2012 Oct 26 &  2.4 & 90    \\  
SDSS J2222+2745 &159 & MMT  & BCS  & 300 line/mm & 2011 Sep 29 &  2.7 & 25    \\  
$\ldots$ & $\ldots$ & MMT  & RCS  & 300 line/mm & 2012 Aug 23 &  2.7 & 25.8,103    \\  
$\ldots$ & $\ldots$ & MMT  & BCS  & 300 line/mm & 2012 Oct 19 &  2.7 & 25   \\  
SDSS J2300+2213 &111 &  MMT  & BCS  & 300 line/mm & 2011 Sep 29 & 1.8   &   45  \\  
\ldots & \ldots &  MMT  & RCS  & 270 line/mm & 2012 Nov 22 &  3.6   &  45   \\  
\ldots  & \ldots & MMT  & BCS  & 300 line/mm & 2012 Dec 12 &  1.8 & 67    \\  
\ldots & \ldots &  MMT  & RCS  & 270 line/mm & 2013 Jan 21 &   1.8  &  50    \\  

\hline
\end{tabular}
\caption{Summary of new spectroscopic observations of SDSS/CASSOWARY lensed galaxies.  We have 
obtained redshifts for all but one of these sources.   We list contaminants that we identified in Table 4.   
The instrument BCS corresponds to the Blue Channel Spectrograph, and RCS refers to the Red Channel 
Spectrograph.   More details are provided in \S2.}
\end{table*}

\subsection{LBT}
  
We obtained near-infrared spectra for an arc toward SDSS J1958+5950, one of the brightest new 
CASSOWARY systems (Table 1), 
using the LUCI near-infrared spectrograph on the Large Binocular Telescope.    We used the 200\_H+K 
grating in two settings, one focused on the z and J-bands (0.95-1.40 $\mu$m) and a second targeting the 
the H and K-bands (1.50-2.30$\mu$m).    We used a longslit with a width of 1\farcs0, providing a reasonably 
low spectral resolving power (R=940-1300).  Data were reduced using an adapted version 
(see Bian et al. 2010 for details) of a near-IR spectral reduction pipeline (G. Becker 2012, 
private communication).   
 
\begin{figure*}
\begin{center}
\includegraphics[width=0.85\textwidth]{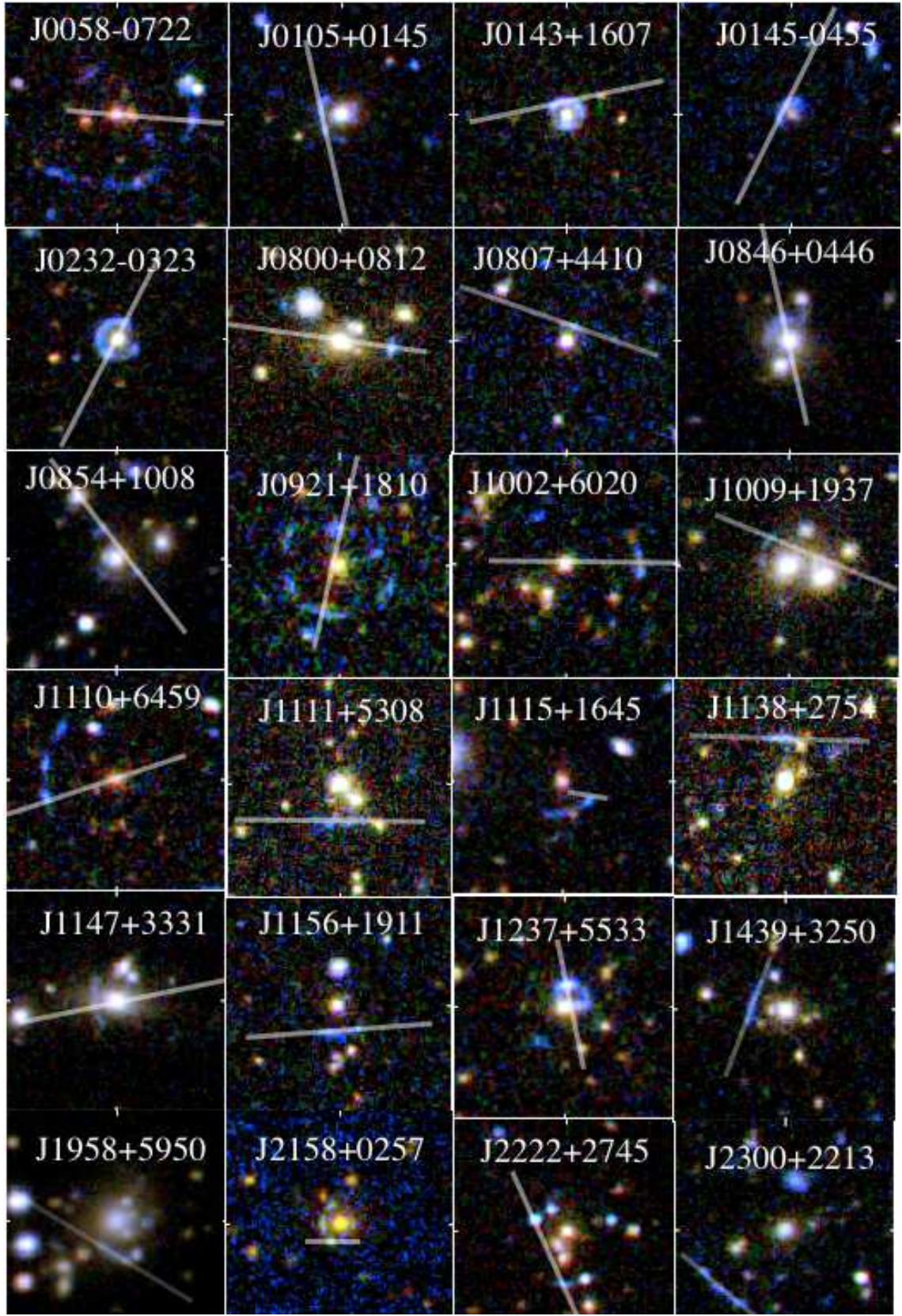}
\caption{Mosaic of arc+lens systems confirmed in this paper.   Each lensed galaxy (except for the arc behind SDSS J1439+3250) has a spectroscopic 
redshift determined from MMT, Magellan, or Keck.   The SDSS IDs are listed at the top of 
each postage stamp.   The images are created from 
gri-band imaging from SDSS and are 40 arcsec on each side.  The slit position used for 
redshift confirmation is overlaid.   For clarity, we do not overlay all slit positions observed in 
cases where multiple position angles or slit centres were taken.}
\label{fig:mosaic}
\end{center}
\end{figure*}

\section{New Spectroscopic Sample}  
  
 \begin{table*}
\begin{tabular}{lllcccccllc}
\hline  Lens Name &  CSWA-ID & z$_{\rm{source}}$ & z$_{\rm{lens}}$ & RA$\rm{_{S}}$ & DEC$\rm{_{S}}$ & RA$\rm{_{L}}$ & DEC$\rm{_{L}}$    & Notes \\  \hline 

SDSS J0058$-$0722 &102 & 1.873 & 0.639$^\dagger$ & 00:58:47.96 & $-$07:21:58.1 &  00:58:48.94 & $-$07:21:56.7 & Abs \\  
SDSS J0105+0145 & 165 & 2.127 &  0.361$^\dagger$ & 01:05:19.89 & +01:44:59.1 & 01:05:19.66  &  +01:44:56.4   & Ly$\alpha$, Abs\\  
SDSS J0143+1607  &116 &  1.499   & 0.415$^\dagger$ &  01:43:50.25 & +16:07:41.0 &  01:43:50.12 & +16:07:39.0 & Abs, Opt  \\  
SDSS J0145$-$0455 &103 & 1.958 & 0.633$^\dagger$   & 01:45:04.18  & $-$04:55:42.7 &  01:45:04.29 & $-$04:55:51.6  & Ly$\alpha$, Abs \\  
SDSS J0232$-$0323 & 164 & 2.518 &  0.450$^\dagger$ & 02:32:49.97 & $-$03:23:29.3 &  02:32:49.86 & $-$03:23:26.0 & Ly$\alpha$, Abs, Opt  \\  
SDSS J0800+0812 &11  & 1.408    &  0.314   & 08:00:12.37 & +08:12:07.0   &08:00:13.06 &   +08:12:08.4   & [OII], Abs  \\  
SDSS J0807+4410 &139 &  2.536 &  0.449 &  08:07:31.37 & +44:10:51.3 &  08:07:31.5 & +44:10:48.5 &  Abs \\  
SDSS J0846+0446 &141 &    1.405  &0.241 & 08:46:47.53 & +04:46:09.3 & 08:46:47.5 & +04:46:05.1 & Opt \\  
SDSS J0854+1008 &142a &    1.437 & 0.298 	&  08:54:28.63  & +10:08:11.3 &   08:54:28.72 & +10:08:14.7  &  [OII] \\  
SDSS J0854+1008 &142b & 1.271 & 0.298 	& 08:54:28.97 & +10:08:19.9  &     08:54:28.72 & +10:08:14.7  & [OII]  \\  
SDSS J0921+1810 & 31 &  1.487 &  0.683 &  09:21:26.46 & +18:10:13.7 & 09:21:25.74 & +18:10:17.3 & [OII]	 \\
SDSS J1002+6020 &117 &  1.114 & 0.571$^\dagger$ & 10:02:00.49 & +60:20:26.5 &   10:02:02.51 & +60:20:26.3  & [OII]  \\  
SDSS J1009+1937 &15 & 2.162 & 0.306 & 10:09:00.06 & +19:37:23.4  &  10:08:59.77 & +19:37:17.5  & Ly$\alpha$ \\  
SDSS J1110+6459 &104 & 2.481  &  0.659$^\dagger$  &  	11:10:19.97 & +64:59:44.7 &11:10:17.69 & +64:59:48.2 & Ly$\alpha$, Abs \\  
SDSS J1111+5308 &16 & 1.945  &  0.412$^\dagger$ &  11:11:03.93  & +53:08:47.1 &  11:11:03.30 & +53:08:51.8 & Abs  \\  
SDSS J1115+1645 &105 & 1.718 & 0.537$^\dagger$ & 11:15:04.02  &  +16:45:34.5 & 11:15:04.39 & +16:45:38.6  & Opt, Abs  \\  
SDSS J1138+2754  &17 & 0.909 & 0.447$^\dagger$ & 11:38:09.00  & +27:54:39.0 &    11:38:08.95 & +27:54:30.7  & [OII], Abs   \\  
SDSS J1147+3331 &107 & 1.205 & 0.212 & 11:47:23.68   & +33:31:53.6  & 11:47:23.30 & +33:31:53.6 & [OII]    \\  
SDSS J1156+1911 &108 &  1.535 &  0.543$^\dagger$ &  11:56:05.5 & +19:11:07.3 & 11:56:05.5  &  +19:11:12.7 & Abs \\  
SDSS J1237+5533 &13 &  1.864 &  0.410 & 12:37:36.13 & +55:33:47.6 & 12:37:36.20 &  +55:33:42.9 & Abs \\  
SDSS J1958+5950 &128 & 2.225  & 0.214$^\dagger$  & 19:58:35.65 &  +59:50:53.6  &  19:58:35.32 & +59:50:58.9  & Abs, Opt \\  
SDSS J2158+0257  &163 &  2.081 &   0.291 & 21:58:43.80  & +02:57:26.8  &   21:58:43.67 & +02:57:30.2  &  Abs, Opt \\  
SDSS J2222+2745 &159a & 2.309   & 0.485 & 22:22:08.58  &  +27:45:24.7 &    22:22:08.57 & +27:45:35.6  & Ly$\alpha$, Abs  \\  
SDSS J2222+2745 &159b &  2.807 & 0.485  &  22:22:09.04 & +27:45:37.6   &  22:22:08.57 & +27:45:35.6  & QSO \\  
SDSS J2300+2213 &111 &  1.93 &  0.443$^\dagger$ &  23:00:18.27 & +22:13:18.0  & 23:00:17.25 & +22:13:29.7  & Abs \\  

\hline
\end{tabular}
\caption{Lensed galaxy redshifts in the CASSOWARY catalog obtained through our spectroscopic campaign.   
Coordinates of the lensed source ($\rm{RA_s}$ and 
$\rm{DEC_s}$) correspond to 
rough location where the slit intersects the arc.    The lens coordinates ($\rm{RA_L}$ and 
$\rm{DEC_L}$) reflect the central lensing galaxy.   In the far right column, we list the methods of redshift confirmation, where 
"Abs" implies rest-UV absorption lines, "Opt" implies rest-optical emission lines.   Those lensed redshifts with a '$\dagger$' 
superscript are not in the SDSS database and are confirmed uniquely via the spectra presented in this paper.   The lens redshift 
for SDSSJ2158+0257 ($z=0.291$) is provided from VLT observations that will be discussed in Deason et al. (2013, in preparation). }
\end{table*}

Through the spectroscopic initiatives presented in \S2, we have confirmed 
the redshifts for 25 bright gravitationally-lensed sources detected in SDSS.  We also measure 
the redshifts of the lensing galaxy (or group) in systems where SDSS spectroscopy was not available.   
With the exception of a lensed arc and quasar 
recently reported  in Dahle et al. 2012 and the $z=0.909$ arc in Wuyts et al. (2012),  these 
redshifts have not appeared in the literature previously.  
Details of the new sample are provided in Tables 1 and 2.   Below we first provide a general overview of the 
redshift confirmation process (\S3.1) before briefly discussing the spectroscopic properties of each system in \S3.2.

\subsection{Redshift Confirmation}

Redshifts were derived through visual examination of the 2D and 1D spectra of each source.   We searched the 
spectra for both emission lines (Ly$\alpha$ or [OII]) and absorption lines.  Following the relations presented in 
Steidel et al. (2010), small offsets were applied to the measured 
redshifts to account for average kinematic offsets of the interstellar absorption lines and Ly$\alpha$.   If possible, the 
values presented in Table 2 include these corrections.  
  In this paper, we focus primarily on  
those systems for which we were able to determine a reliable redshift.   At the end of this section, we also briefly 
discuss interloping objects which satisfied our search algorithm.  

The majority of the new  lensed sources have been confirmed with the MMT 
Blue Channel Spectrograph.   When bright stellar continua is strongly detected in our spectra, redshift confirmation is easily 
achieved through identification of prominent far-UV absorption lines and Ly$\alpha$ emission or absorption (Figure 2).  
The presence of strong Ly$\alpha$ emission enables redshift confirmation in galaxies with weaker continua.  
As the sensitivity of the MMT blue channel spectra degrades at wavelengths shorter than 3800~\AA, each of the
five sources confirmed primarily via detection of Ly$\alpha$ emission have redshifts of $z>2.1$.   In cases where 
Ly$\alpha$ emission is the primary method of redshift identification, we additionally identify weak interstellar absorption lines, thereby confirming the interpretation of the emission line as 
Ly$\alpha$.   

Reliable redshift identification at $z\lsim 1.7$ is more difficult with MMT Blue Channel spectroscopy as many 
of the strongest absorption lines are no longer visible in the spectral window.   In this redshift range, most 
redshifts have been confirmed through detection of [OII] emission with either the MMT Red Channel 
Spectrograph or Keck/DEIMOS.    Of the sample in Table 2, 7 sources were initially confirmed via [OII] emission 
with redshifts spanning $0.91<z<1.49$.    For those sources observed with MMT, we typically also identified 
absorption from Mg II, Fe II, or Al II transitions.   For systems 
observed with DEIMOS, our chosen grating was able to resolve the [OII] doublet, enabling definitive redshift 
confirmation.   For several galaxies, redshifts were first 
obtained via infrared spectroscopy.   In these cases, optical spectra were taken 
prior to infrared observation, yet in spite of the detection of multiple absorption lines, we 
were unable to derive a confident redshift.   Infrared  spectroscopy yielded detection 
of multiple rest-optical emission lines which enabled confident redshift identification.  

\begin{figure*}
\begin{center}
\includegraphics[width=0.47 \textwidth]{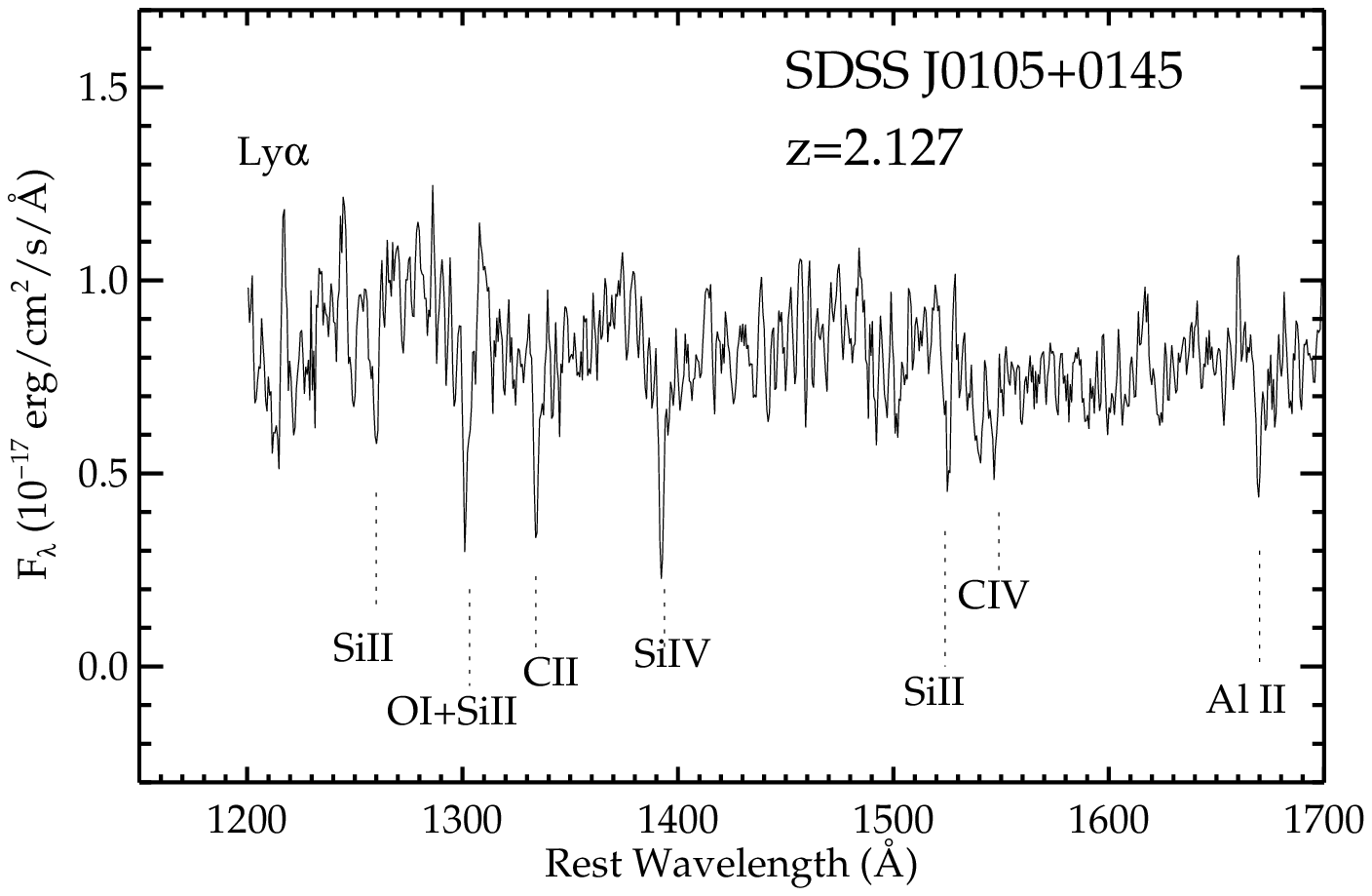}
\includegraphics[width=0.47 \textwidth]{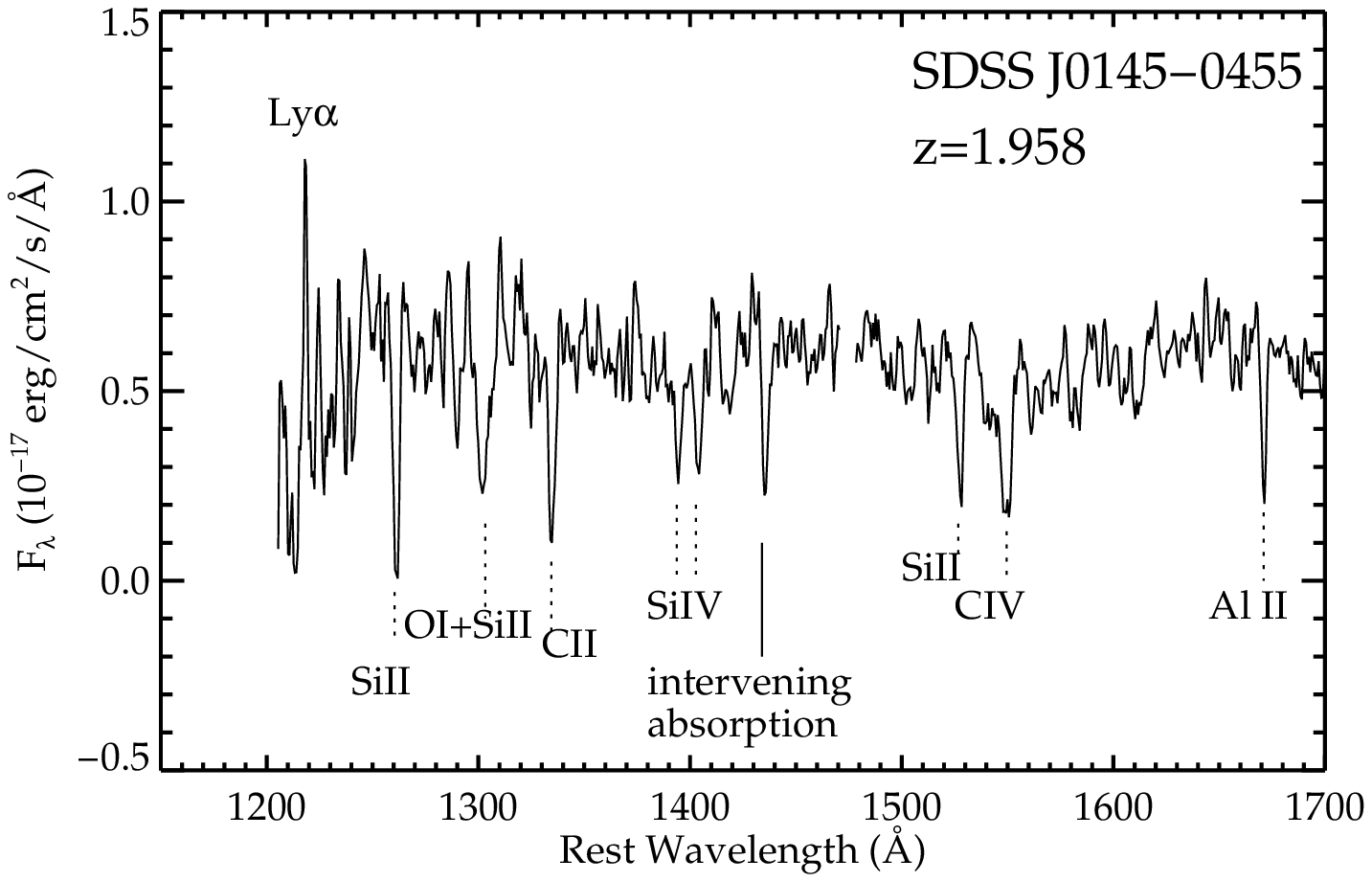}
\includegraphics[width=0.47 \textwidth]{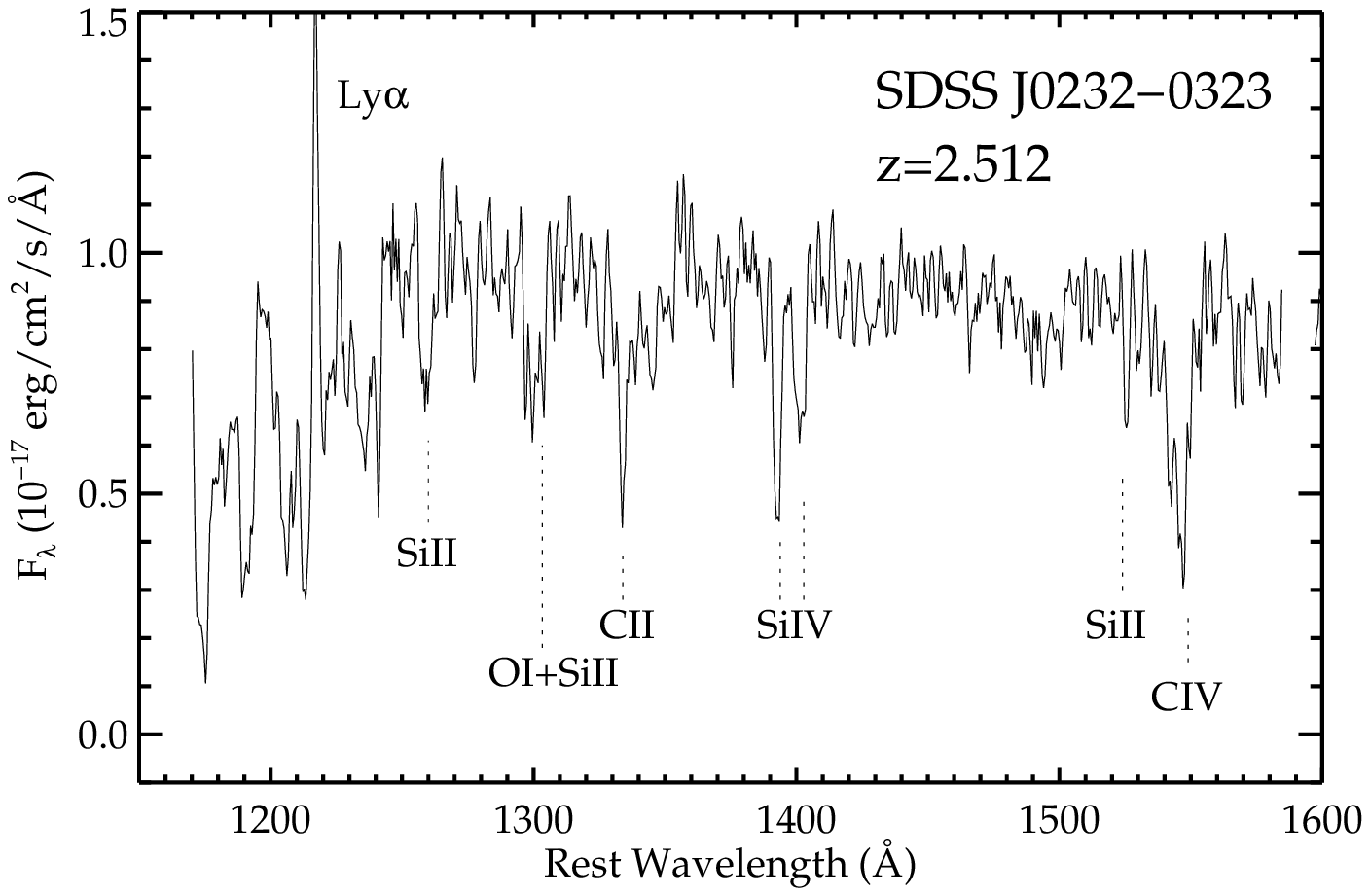}
\includegraphics[width=0.47 \textwidth]{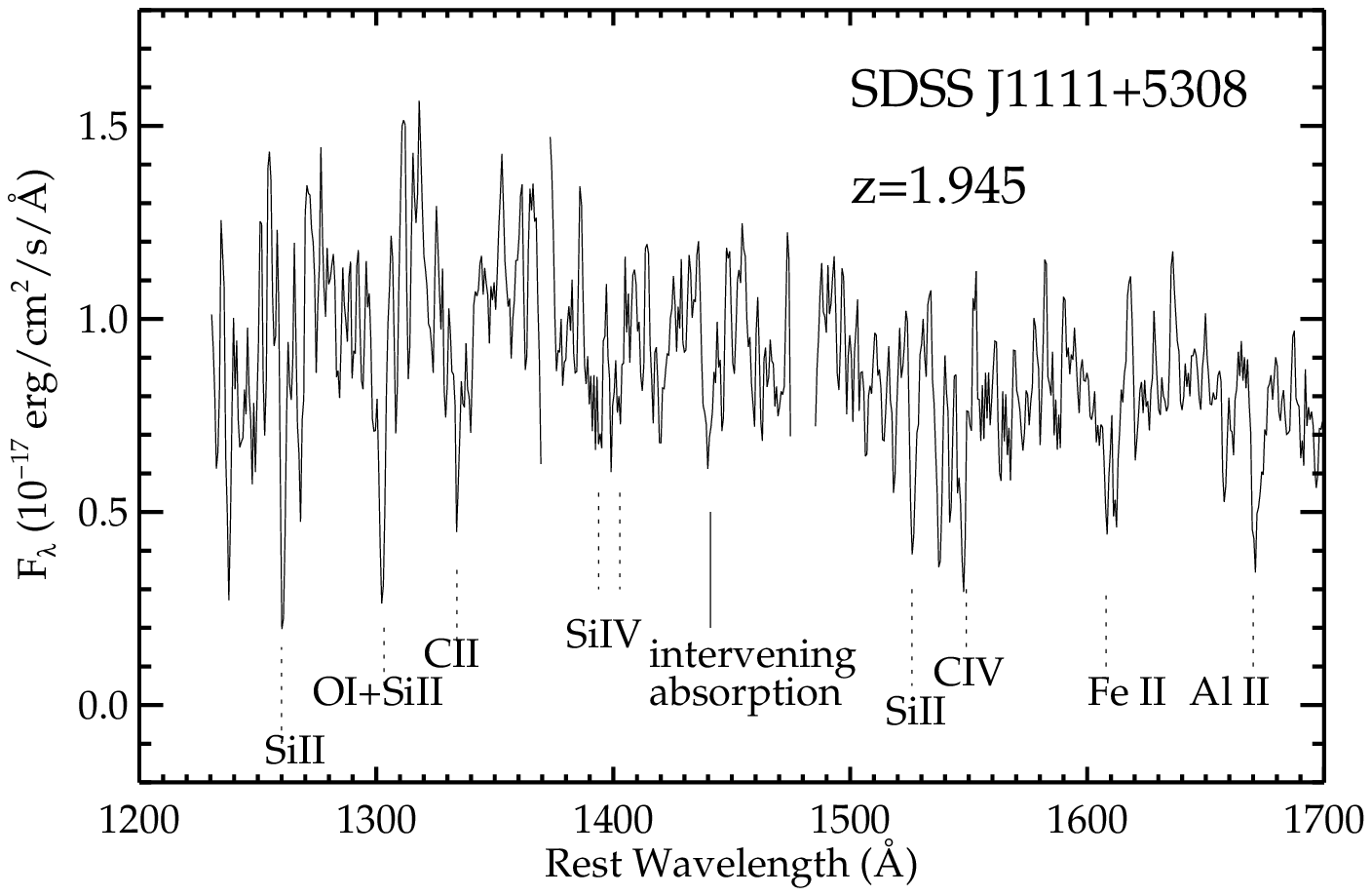}
\includegraphics[width=0.47 \textwidth]{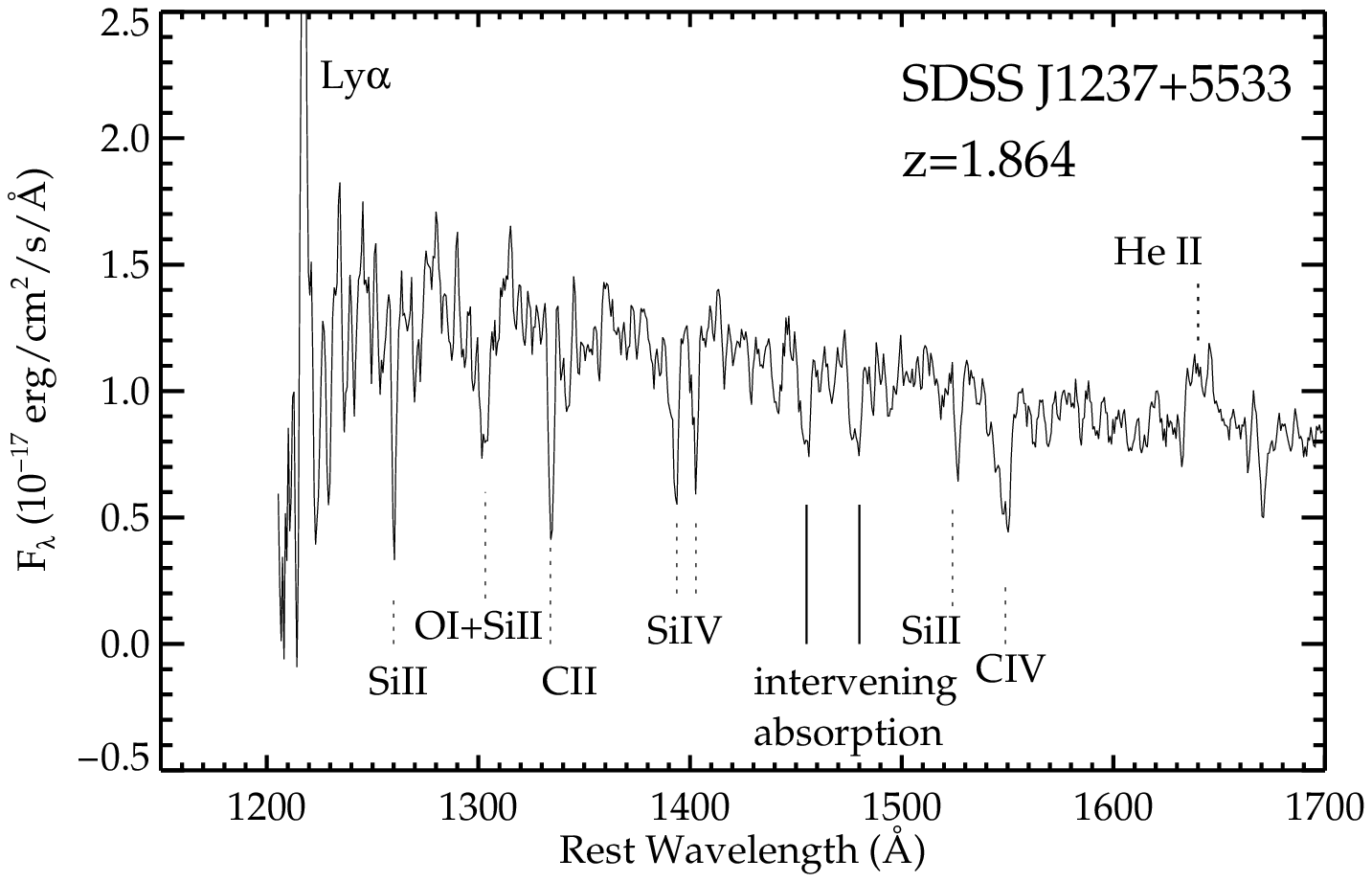}
\includegraphics[width=0.47 \textwidth]{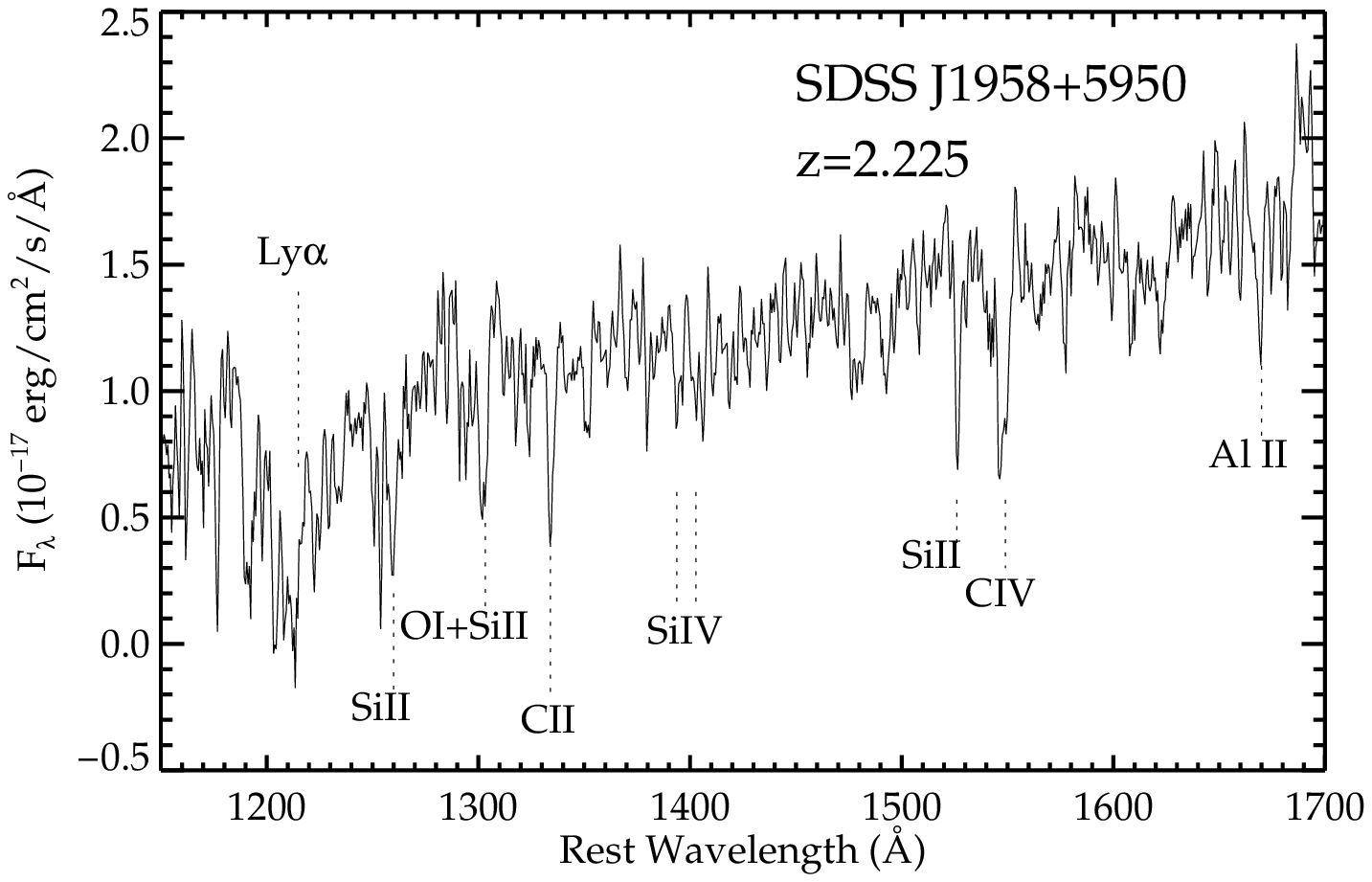}
\caption{Examples of MMT spectra of newly-confirmed gravitationally lensed galaxies located within SDSS as 
part of the CASSOWARY survey.   Owing to the very bright optical magnitudes ($r\simeq 19-22$), exposure times 
of just 30-45 min with the Blue Channel Spectrograph (see Table 1) are sufficient to reveal many absorption 
lines (denoted with dotted lines) throughout the rest-frame UV.  We also point out several likely intervening 
absorption systems identified in the 2D and 1D spectra.     }
\label{fig:absspectra}
\end{center}
\end{figure*}

Lens redshifts are of course required to derive the intrinsic properties of the lensed source.  While many of the lens candidates 
have redshifts from SDSS spectroscopy, there are a number of early-type systems in our database without 
spectra.  As part of our program with MMT, we obtained spectra of 13 early-type lensing galaxies 
lacking  redshifts.    We detected strong continuum and absorption features in each system, allowing us to 
accurately derive the lens redshifts.   These lenses are marked in Table 2.   

Throughout our spectroscopic campaign, we also identified a small number of contaminants in our 
galaxy-galaxy lens selection.   Most of the contaminant systems instead correspond to blue galaxies 
at the same redshift as the red galaxy we had identified as a candidate lens.  The coordinates  
of those systems we confirmed as contaminants are listed in Table 3.  

\subsection{Notes on Individual Galaxies}

Below we provide brief descriptions of the spectra of the galaxies confirmed as part of this survey.   When 
possible, we state redshifts of absorption lines, Ly$\alpha$ emission, as well as the systemic redshifts 
derived from emission lines in near-IR spectroscopy.  

{\bf SDSS J0058$-$0722, z=1.87:}   This system is composed of a number of blue arcs around three bright early type galaxies 
at wide separation ($\simeq 15$\arcsec).    We detected the brightest source continuum flux from the arc to the west of 
the lenses.   Redshift confirmation is achieved through identification of absorption lines (Si IV, Si II, CIV, Fe II, Al II) in the 
MMT blue channel spectrum.    We determined the redshift of the absorbing gas (z=1.872) from Al II, the most isolated 
absorption line in the blue channel spectrum.     The red channel spectrum shows absorption at the observed 
wavelengths expected for Fe II $\lambda$2344 and Fe II $\lambda\lambda$2374,2382.     As the lens redshift 
is not in the SDSS database, we have obtained additional exposures with the MMT slit oriented along the three central 
group members.    We measure a lens redshift of $z=0.639$ through identification of absorption lines in the MMT 
spectrum.

{\bf SDSS J0105+0145, z=2.13. }   Strong stellar continuum and prominent absorption lines are seen in the 
MMT blue channel spectrum of the arc behind SDSS J0105+0145 (Figure 2).    
We measure a redshift of $z=2.126$ from CII $\lambda$1334 absorption.   
Weak Ly$\alpha$ emission is tentatively detected with $z_{\rm{Ly\alpha}}=2.131$.
The mean kinematic offset between the blueshifted absorbing gas and Ly$\alpha$ emission is 540 km s$^{-1}$.   
Higher S/N data will be required to confirm the properties of Ly$\alpha$ emission.  We note that the 
apparent strength of SIV absorption is artificially amplified by a negative skyline residual.   We measure the 
lens redshift ($z=0.361$) with a separate MMT exposure.

{\bf SDSS J0143+1607, z=1.50.}  Strong continuum is detected in the MMT blue channel spectrum, but some blending with the 
lens ($\simeq 2$ arcseconds away from the centre of the slit) is present.   Nevertheless we detect CIV$\lambda$1549 
in absorption as well as weak absorption corresponding to Al II $\lambda$1671.  The CIV absorption line redshift 
in the MMT spectrum is $z=1.500$.    We detect [OII] emission from two components of the source in our MMT 
red channel spectrum.   The line is unresolved at the resolution of our spectrum.   Assuming a flux-weighted 
mean [OII] wavelength of 3728.3~\AA, we derive a redshift of $z=1.500$ from the line detection.   
As the central lens galaxy redshift was not in the SDSS database, we  obtained its redshift ($z=0.415$) with  an 
MMT red channel spectrum.

{\bf SDSS J0145-0455, z=1.96:} We detect strong continuum and absorption in the spectrum of SDSS J0145-0455 (Figure 
2).   We measure a mean low ionisation absorption line redshift of $z=1.957$.   Weak Ly$\alpha$ emission is detected, 
but the rest-frame EW ($\sim 6$ \AA) is fairly uncertain owing to the difficulty in characterising the 
continuum at wavelengths shorter than  3800~\AA.  
CIII]$\lambda$1909 emission is detected with a measured rest-frame EW of 1.5~\AA,  comparable 
to that seen in the composite $z\simeq 3$ LBG spectrum of Shapley et al. (2003).

We detect pronounced absorption from an intervening system at observed wavelength 4244 ~\AA\ (corresponding 
to rest-wavelength 1435~\AA\ in the frame of the lensed $z=1.96$ galaxy).   We cannot confidently determine 
the redshift of the absorption system with our current low resolution spectra.   If due to Mg II absorption, it would imply a redshift of 
$z=0.52$ and might indicate some spectral blending between Fe II $\lambda$2600 absorption (from the intervening system) and 
CII$\lambda$1334 absorption from the $z=1.96$ source.   Likewise, if the absorption at 4244~\AA\ is due to CIV at $z=1.74$, 
then it might signal spectral blending between Al II $\lambda$1671 (from the intervening system) and 
CIV$\lambda$1549 in the $z=1.96$ reference frame.   We do note the presence of weak absorption at 4188~\AA, as might be expected from 
Si II$\lambda$1526 at $z=1.74$.     The finally possibility we consider is that the absorption at 4244~\AA\ corresponds to 
Fe II$\lambda$2600 at $z=0.63$.    This possibility is particularly intriguing, as this redshift is identical to the redshift 
of the galaxy responsible for lensing, potentially signalling some residual low ionisation gas associated with the early-type system.   
In this case, we would expect spectral blending from Mg II absorption (at $z=0.63$) at the wavelength of CIV in the $z=1.96$ 
reference frame.     Higher resolution spectra will help clarify the situation. 
 
{\bf SDSS J0232$-$0323, z=2.51} As seen in Figure 1, the MMT 
slit targeting SDSS J0232$-$0323 covers the source in two 
locations, to the NW and SE of the lens.  We detect Ly$\alpha$ emission and 
interstellar absorption lines from the source on both sides (Figure 2).      The escaping Ly$\alpha$ radiation has a 
mean redshift of $z_{\rm{Ly\alpha}}$=2.515.   From the unblended CII$\lambda$1334 absorption line, we 
find that the low ionisation absorbing gas has a mean redshift of $z=2.510$.  We establish a systemic 
nebular redshift of $z=2.512$ from detection of [OIII]$\lambda$5007 in our Magellan/FIRE spectrum.
These data imply that the low ionisation gas has a mean outflow velocity of -160 km s$^{-1}$ and 
that the Ly$\alpha$ radiation escapes with a mean velocity of 250 km s$^{-1}$ with respect to systemic.  

{\bf SDSS J0800+0812, z=1.41:}  Both blue and red channel data were obtained for SDSS J0800+0812, 
providing spectral coverage between 3600 and 9070~\AA.    Absorption from Fe II$\lambda$2600 and 
Mg II$\lambda\lambda$2796,2803
is seen in the MMT red channel spectra, indicating a redshift of $z=1.407$ for the absorbing gas.   Additionally, we 
detect strong [OII] emission (rest-frame EW of 60~\AA) at 8983~\AA,  revealing a nebular redshift of $z=1.408$ assuming 
that the rest-frame centroid of the unresolved [OII] doublet is 3728.3~\AA.    

{\bf SDSS J0807+4410, z=2.54:}   We detect weak stellar continuum, Ly$\alpha$ absorption, and multiple 
interstellar absorption lines in our MMT blue channel spectrum of the faint arc around SDSSJ0807+4410.   The low 
ionisation absorbing gas has a mean redshift of $z=2.540$ in the MMT spectra.  
Its close proximity to the lens ($\simeq 3$ arcsec to the centre of the elliptical galaxy) results in some 
blending at wavelengths greater than 5500~\AA.  

{\bf SDSS J0846+0446, z=1.43:}   Redshift confirmation was first achieved through the detection of numerous 
emission lines with DEIMOS.    In addition to [OII] and [Ne III]$\lambda$3869, we identified Hydrogen 
Balmer series lines  H$\epsilon$, H8, H9, H10, H11, H12.    Subsequent 
infrared observations with Magellan/FIRE yielded many further rest-optical emission features, including the 
temperature-sensitive [OIII]$\lambda$4363 auroral line (Figure 3) and the electron density sensitive [SII] emission 
lines.   We derived a nebular redshift of $z=1.425$ from the observed hydrogen Balmer emission lines in the 
FIRE spectrum.    The observed spectrum points to emission from a very young and metal-poor system.  
The detection of so many emission lines will enable a comprehensive analysis of the physical conditions of the 
ionised gas.   We will briefly discuss this source in more detail in \S5.1.   A complete discussion will be 
presented in a future paper (Stark et al. 2013, in preparation).   

{\bf SDSS J0854+1008, z=1.44 and z=1.27.}  DEIMOS observations of the SDSS J0854+1008 system yielded 
detections of [OII] emission at two different redshifts.   We detect resolved [OII] emission from the blue 
arc to the NE of the lens (see Figure 1), yielding a nebular redshift of $z=1.271$.  
We also detect faint emission at  3868~\AA\ that we tentatively identify as [Ne III]$\lambda$3869.
In addition to this source, the DEIMOS slit also passes through a fainter ring of blue emission 
surrounding the lens.  The slit intersects the ring to the NE and SW of the lens.   We detect 
resolved [OII] emission at 9080, 9086~\AA\ in both images, confirming the redshift of this second source 
as $z=1.436$.   

{\bf SDSS J0921+1810  z=1.49:}  Redshift confirmation is achieved through detection of [OII] emission in both the 
DEIMOS and MMT red channel spectra.   The doublet is fully resolved in the 1200 line/mm grating 
DEIMOS spectrum.     We derive a nebular redshift of $z=1.486$ from the centroids of the [OII] doublet.   We 
do detect stellar continuum in the DEIMOS spectrum, but the narrow wavelength range does not cover strong 
absorption features.   The shallower MMT spectrum does not reveal continuum, so at this stage, we 
have no information on the properties of the circumgalactic gas around 
SDSS J0921+1810.    
 
\begin{figure*}
\begin{center}
\includegraphics[width=0.47 \textwidth]{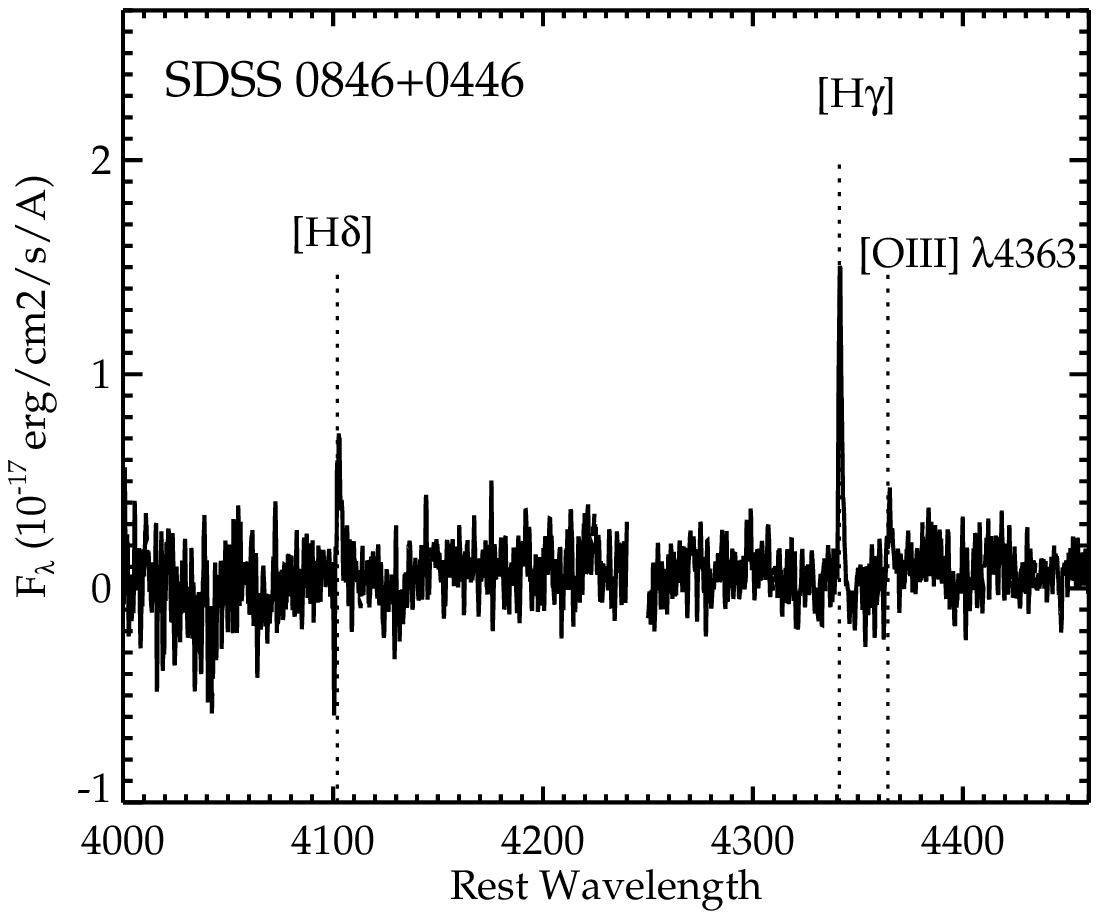}
\includegraphics[width=0.47 \textwidth]{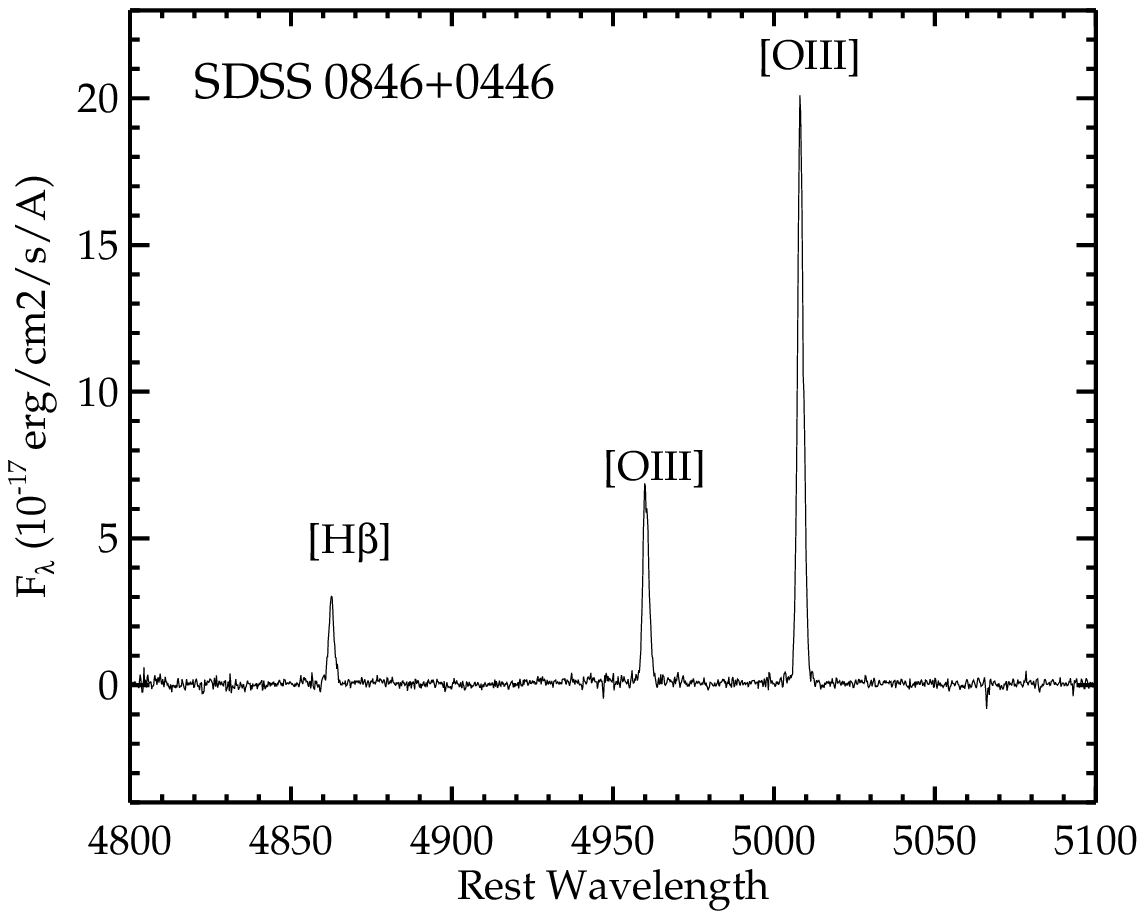}
\caption{Portion of Magellan/FIRE near-IR spectra of  SDSS J0846+0446, a lensed strong emission line $z=1.43$ galaxy 
in our new sample.    Detection of the [OIII]$\lambda$4363 emission 
line (left panel) enables measurement of the electron temperature and reflects the low metallicity (0.2 Z$_\odot$) 
of the nebular gas.  The rest-frame EW of [OIII]$\lambda$5007 ($\gsim 320$~\AA; right panel) is similar to those of extreme line emitting 
galaxies found in grism and imaging surveys, while the EW of Hydrogen Balmer lines points to a young stellar population.  
With numerous emission lines detected across the optical and near-IR, future analysis of 
this source will yield new insight into the nature of the extreme line emitting population.      }
\label{fig:absspectra}
\end{center}
\end{figure*}

{\bf SDSS J1002+6020, z=1.11:}   We detect  [OII] emission in each of the three spectra taken of the arc lensed 
by the galaxy SDSS J1002+6020.   Based on the wavelength of [OII] emission, we confirm a redshift of $z=1.114$ for the 
source.     We detect emission at the wavelength of H$\delta$ and Ne III $\lambda$3869 in the MMT red channel 
spectrum.   The MMT blue channel spectrum shows absorption at the observed wavelengths corresponding to 
Fe II $\lambda$2344 and  Fe II $\lambda\lambda$2374, 2382.

{\bf SDSS J1009+1937, z=2.16:}   Ly$\alpha$ emission is present at 3850~\AA\ in the MMT blue channel spectrum, 
indicating a Ly$\alpha$ redshift of  2.167.   While the continuum S/N is not as strong as in other sources, we confirm 
the presence of the Si II$\lambda$1526, and CIV$\lambda$1549 absorption lines.    The CIV centroid is consistent 
with a redshift $z=2.158$.  Higher S/N absorption line spectra are required to accurately quantify the kinematics of 
the absorbing gas.  

{\bf SDSS J1110+6459, z=2.48:}   Ly$\alpha$ emission is clearly visible in  
the MMT spectrum of SDSS J1110+6459.  The equivalent width is moderate in strength (rest-frame $EW\simeq 3$~\AA) 
with a redshift of $z=2.483$.    We detect absorption from OI+SiII$\lambda$1303, CII$\lambda$1334, and 
Al II $\lambda$1671 with a mean redshift of $z=2.476$.      

{\bf SDSS J1111+5308, z=1.95:}  Strong continuum and absorption are detected in the MMT blue channel 
spectrum, as is seen in Figure 2.    The mean redshift for the low ionisation absorbing 
gas is $z=1.946$.   We detect absorption at 4245~\AA\ which would correspond to 1441~\AA\ if in the 
rest-frame of the lensed source toward SDSS J1111+5308.   As no absorption is expected at this rest-wavelength, 
this feature is likely produced by gas associated with a foreground system.   Higher resolution 
data will be required to determine the redshift of the intervening system.  

{\bf SDSS J1115+1645, z=1.72:}   Our initial MMT optical spectra of this target suffered  contamination from a 
foreground red galaxy (see postage stamp Figure 1), making redshift confirmation difficult.   After placing the FIRE slit directly 
on the bright blue knot to the SW of the lens, we were able to confirm the systemic redshift as $z=1.717$ through 
detection of H$\alpha$, H$\beta$, and [OII] emission.   Emission from [OIII]$\lambda$5007 and 4959 falls in the low transmission region 
between the J and H-bands and 
is not detected.   No redshift existed for the lens in the SDSS database.    After redshift confirmation with FIRE, we therefore revisited 
the system with MMT to acquire the lens redshift.   By orienting an MMT red channel slit to include both source and lens, we were 
able to identify lens absorption features indicating a redshift of $z=0.537$.     We also identified continuum from the $z=1.72$ source, but the 
S/N was not sufficient to reliably characterize Mg II and Fe II absorption lines.    We note that observations of 
SDSS J1115+1645 in Bayliss (2012) reveal an additional lensed source at $z=3.463$.   

{\bf SDSS J1138+2754  z=0.91:} Extended [OII] emission is detected in both blue and red channel spectra, revealing   
the source redshift as $z=0.909$, as previously reported in Wutys et al. (2012).   We measure a rest-frame [OII] EW of 60~\AA, similar to SDSS J0800+0812.  
We also detect extended [Ne III] in the red channel spectrum.    We do not detect strong absorption from Fe II or Mg II 
transitions.

{\bf SDSS J1147+3331, z=1.21:}    Redshift confirmation ($z=1.205$) was obtained through detection of [OII] emission 
with our MMT red channel spectrum.    The brightness of the lens  ($z=16.1$) and its reasonably close proximity
 to the source (see Figure 1) cause significant blending in the MMT spectrum.   While we detect 
faint stellar continuum (particularly in the blue channel data), it is difficult to reliably characterise absorption lines as a 
result of the blending.     

{\bf SDSS J1156+1911, z=1.54:}  Strong continuum in the MMT blue channel spectrum.  We detect absorption features 
at $\sim 3870$, 3927, 4076, and 4236~\AA.   The most likely classification of these features is Si II $\lambda$1526, CIV $\lambda$1549, Fe II $\lambda$1608, and Al I I$\lambda$1671, respectively.   The centroids of these absorption features imply a redshift of $z=1.535$ for the absorbing gas.  In the same spectrum, we note the presence of weak low S/N emission at 4226 and 4839~\AA\ (OIII] $\lambda$1667~\AA and CIII] $\lambda$1909~\AA) and 
weak absorption at 5940~\AA possibly corresponding to Fe II $\lambda$2344.  The S/N of the red channel is lower than 
that of our blue channel spectrum, but we do tentatively note absorption at 7090-7080~\AA, as expected from the Mg II doublet.

{\bf SDSS J1237+5533, z=1.87:}  In addition to being among the brightest lensed galaxies in our sample, 
the arc toward SDSS J1237+5533 is fairly compact, translating into a very high surface brightness and easily detectable 
continuum emission.   Strong absorption lines are readily apparent across the 
rest-UV (Figure 2).    We obtain a redshift 
of $z=1.863$ from CII $\lambda$1334 and $z=1.868$ from 
Ly$\alpha~\lambda$1216.    The EW of Ly$\alpha$ emission is uncertain due to the weak continuum signal 
at 3400~\AA.  The velocity offset between Ly$\alpha$ emission and CII absorption is 570 km s$^{-1}$, consistent 
with expectations from spectra at $z\simeq 2-4$ (Shapley et al. 2003; Steidel et al. 2010; Jones et al. 2012).  
In addition to Ly$\alpha$, we tentatively identify emission from He II$\lambda$1640 and 
CIII]$\lambda$1909.  The rest-frame equivalent width of CIII]$\lambda$1909 (4~\AA) is greater than 
the equivalent width measured in the $z\simeq 3$ composite spectrum but consistent with that seen in the subset 
of galaxies with strong Ly$\alpha$ emission (Shapley et al. 2003).  
We identify intervening absorption at $\lambda=$4167 and 4229~\AA.   Possible 
interpretations for these features include Si II$\lambda$1526 and CIV$\lambda$1549 absorption at $z=1.73$ or 
Fe II$\lambda$2344 and Fe II$\lambda$$\lambda$2374,2382 absorption at $z=0.78$.   
Higher resolution and higher S/N spectroscopy will be required to confirm the redshifts of the intervening 
absorption systems and disentangle their contribution to the absorption associated with the CGM of the $z=1.87$ lensed galaxy.  

\begin{table}
\begin{tabular}{lccc}
\hline  Contaminant Name &  RA & DEC & z$_{\rm{cont}}$ \\ \hline

SDSS J0840+1052 & 08:40:21.0 & +10:52:12.7 & 0.048 \\  
SDSS J0013+3512 &  00:13:31.9 & +35:12:21.0   & 0.144, 0.269 \\  
SDSS J1046+1048 & 10:46:01.1 & +10:48:52.7 & 0.247 \\  
SDSS J0206+0448  & 02:06:38.9 & +04:48:03.8 & 0.262 \\  
SDSS J1502+2920 & 15:02:36.6 & +29:20:53.9 & 0.245 \\  
SDSS J1421+4143 & 14:21:55.3 & +41:43:20.9 & 0.228 \\  
\hline
\end{tabular}
\caption{Contaminants in CASSOWARY sample.   The redshift of the blue 
galaxy we had identified as a candidate arc is listed as z$_{\rm{cont}}$.  Coordinates 
give the location of the red central galaxy.   }
\end{table}

{\bf SDSS J1439+3250:}  We detect continuum and several pronounced absorption lines in the MMT blue channel spectrum 
of the triplet arc system to the southwest of the SDSS J1439+3250 lens.    The redshift identification appears complicated by 
the presence of intervening absorption systems.    We currently identify a number of possible redshift solutions, including 
$z=2.15$ based on absorption at 3970, 4202, 4806, and 4879~\AA\ which would correspond to Si II $\lambda$1260, CII $\lambda$1334, 
SiII $\lambda$1526, and CIV $\lambda$1549.   The red channel spectra (Table 1) do not 
show strong absorption features, perhaps due to limited exposure time and reduced transparency during the 
observations.  We note that the absorption seen at 3970~\AA\ in the blue channel spectra might instead  reflect absorption from 
the Mg II doublet at the redshift of the lens ($z=0.418$).   If confirmed, this would indicate the presence of neutral gas 
associated with the central lens or satellite galaxies.   Higher S/N spectra are required to finalise the redshift of the 
source and the interpretation of the absorption lines.

{\bf SDSS J1958+5950, z=2.22:}  At r=19.6, this is one of the brightest lensed galaxies known in SDSS.   
Multiple strong absorption lines are detected in the MMT spectrum (Figure 2).   We derive a redshift of $z=2.222$ 
from CII $\lambda$1334 absorption, consistent with the redshifts derived from the other low ionisation absorption lines.    While 
the lens is 6 arcsec away from the slit centre, the red continuum spectral slope possibly points to some 
low level contamination from one of the lensing galaxies.    However the presence of strong absorption lines from the 
$z=2.22$ galaxy suggests that contaminating stellar continuum is not dominant in the wavelength regime we 
are considering.     Higher resolution imaging will be required to decipher the UV continuum slope of the components 
of the arc.    As detailed in Table 1, we have also obtained near-IR spectra covering 
0.9 to 2.4$\mu$m with LBT.    Multiple emission lines and continuum are detected in the spectrum.   We derive a mean 
systemic redshift of $z=2.225$ from H$\alpha$ and [OIII]$\lambda$5007 emission.   The low ionisation 
absorption lines detected with MMT are therefore blueshifted by 200-300 km s$^{-1}$ from the systemic redshift.    

{\bf SDSS J2158+0257, z=2.08} Multiple absorption lines are detected throughout the 
MMT blue channel spectrum.    We derive a mean low ionisation absorption line redshift of 
$z=2.079$ from CII $\lambda$1334 and OI+SiII $\lambda$1303.   Ly$\alpha$ is seen in absorption.  
Strong emission lines are present throughout the rest-frame optical with Magellan/FIRE.   
The systemic redshift determined from H$\beta$ and H$\alpha$ is $z=2.081$, indicating that 
the average velocity of the low ionisation absorbing gas is -200 km s$^{-1}$.   The lens redshift ($z=0.290$) is confirmed 
through VLT/FORS spectroscopy (Deason et al. 2013, in preparation).

{\bf SDSS J2222+2745, z=2.30 and z=2.81:} 
We confirmed a lensed 
galaxy at $z=2.31$ and a quasar at $z=2.81$ in the first spectrum of we took of the SDSS J2222+2745 system.   
Both sources were recently reported 
in Dahle et al. (2012).   The galaxy redshift is easily identified via the presence of Ly$\alpha$ 
emission and weak interstellar absorption lines.   The Ly$\alpha$ rest-frame EW is 20~\AA, 
the strongest in our sample.   The low ionisation absorbing gas ($z=2.298$) is blueshifted with 
respect to the escaping Ly$\alpha$ radiation (z$_{\rm{Ly\alpha}}=2.300$), consistent with a 
kinematic offset of 250 km s$^{-1}$.

The QSO toward SDSS J2222+2745 is easily confirmed in our MMT spectra through broad 
Ly$\alpha$ and CIV emission.   Ly$\alpha$ emission is extended over 16~arcseconds.  As 
detailed in Dahle et al. (2012), the lensed quasar is multiply-imaged with 6 separate 
images reported.   Over the 
last year, we have obtained spectra for 4 quasar images with MMT.  The 
strong continuum in the quasar  allow us to characterise the spatial extent of the absorbing gas from 
the nearby $z=2.30$ lensed galaxy.   While this is common in QSO absorption systems, the  
brightness of the arc ($g\simeq 21$; Dahle et al. 2012) enables unique constraints 
on the absorbing gas along the line of sight to the galaxy.   
We will provide a brief discussion of this particular system in \S5.     

\begin{figure*}
\begin{center}
\includegraphics[width=.9 \textwidth]{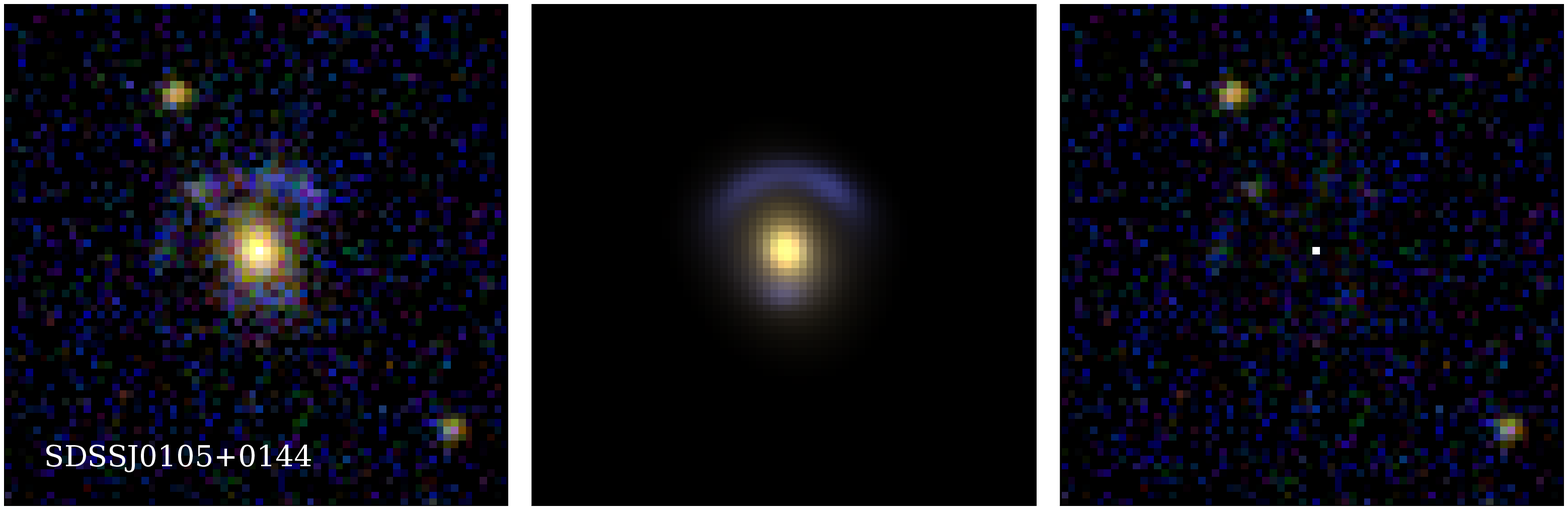}
\includegraphics[width=.9 \textwidth]{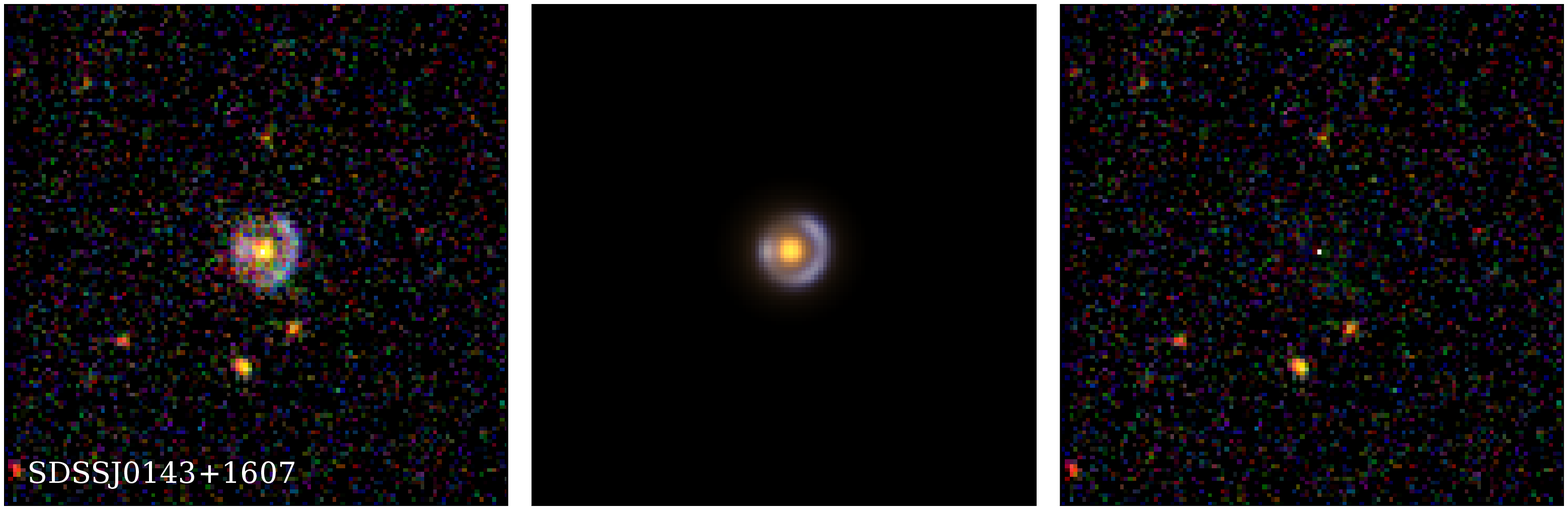}
\includegraphics[width=.9 \textwidth]{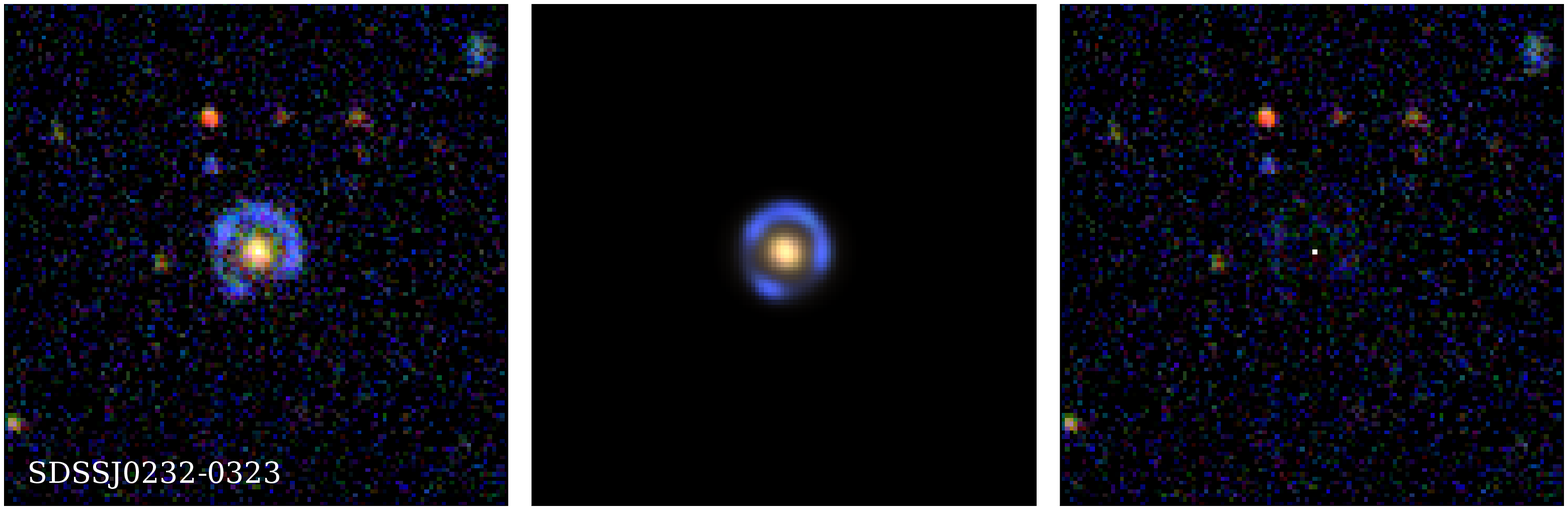}
\includegraphics[width=.9 \textwidth]{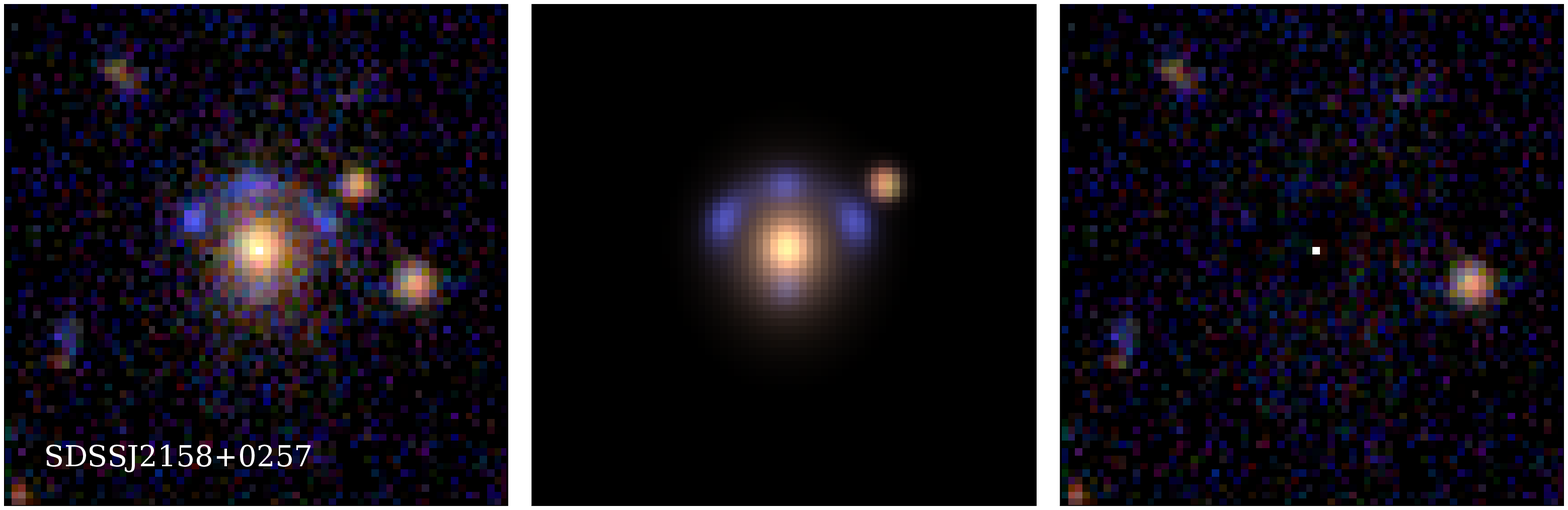}
\caption{Examples of lens models for newly-confirmed CASSOWARY systems that demonstrate the robustness of our modelling scheme.  The left panel shows the SDSS images, the middle
panel shows the model of the system, and the right panel shows the residuals when the model is subtracted from the data.   Details 
of the modelling procedure are provided in \S4.2 and lens model parameters can be found in Table 4.}
\label{fig:models}
\end{center}
\end{figure*}

{\bf SDSS J2300+2213, z=1.93:}  We have observed the arcs in SDSS J2300+2213 on numerous occasions with MMT.   
We identify strong continuum and a number of absorption lines.   In the MMT blue channel data, we identify absorption 
lines at $\simeq  $3816, 3906, 4473, 4543, and 4893~\AA, which appears to correspond to absorption from OI+SiII$\lambda$1303, CII $\lambda$1334, 
Si II $\lambda$1526, and CIV $\lambda$1549. and Al II $\lambda$1671  
absorption at $z=1.93$.   The S/N of the detected absorption lines does not permit a more precise redshift measurement. 
  We also see absorption at $\simeq 4140$~\AA\ which we might stem from an 
intervening absorption systems.    In the same blue channel spectrum, we also see absorption at the observed wavelength 
corresponding to  and Fe II $\lambda\lambda$2374, 2382.   At $z=1.93$, the Fe II $\lambda$2344 and Fe II $\lambda\lambda$2586,2600 transitions 
overlap with the atmospheric B-band and A-band absorption features and so cannot be used to independently verify the redshift.
Using our MMT red channel data, we do note the presence of absorption at $\simeq 8200$~\AA, which would correspond to 
Mg II  doublet at $z=1.93$.   We have determined the redshift of the central lensing galaxy ($z=0.443$) using a combination 
of red channel and blue channel spectra.

\section{Physical Properties of CASSOWARY lensed sources}

We present the total CASSOWARY spectroscopic sample (\S4.1).
Derivation of physical properties of course requires  
estimates of the source magnification which we describe in \S4.2.   Using these 
measurements, we discuss the typical star  formation rates and stellar masses of 
lensed sources in our sample (\S4.3).    

\subsection{Total spectroscopic sample}

We have added spectra for 25 bright objects to our redshift database of lensed sources 
in SDSS.   With the exception of the lensed QSO and galaxy system behind SDSS J2222+2745 
(recently reported in Dahle et al. 2013) and the $z=0.91$ arc toward SDSS 1138+2754 (Wuyts et al. 2012), 
the redshifts are new additions to the literature.   Several of the new sources had  appeared (without 
spectroscopic confirmation) in Wen et al. (2011), and several are mentioned as part of the Sloan 
Giant Arcs Survey (SGAS) in Bayliss (2012).

The total CASSOWARY SDSS spectroscopic catalog (now numbering 55 sources) is listed in Table 4.  
Many objects in this sample are of course not unique to the CASSOWARY selection.    The 
Sloan Bright Arcs Survey (SBAS) has identified (and provided initial confirmation) for a significant 
number of galaxies listed in Table 4 (see reference column in the table) using an algorithm selecting  
blue arcs around both luminous red galaxies and brightest cluster galaxies (e.g., Allam et al. 2007; 
Kubo et al. 2010).   As mentioned above, there is also overlap with the SGAS sample.   Nine of the 
CASSOWARY targets listed in Table 4 were observed in Bayliss et al. (2011), and we expect to 
see more overlap as the entire SGAS sample is released.

\subsection{Typical source magnification factors}

We use the $gri$ imaging data from SDSS DR8 to fit each of the objects with 
singular isothermal ellipsoid 
\citep[SIE;][]{kormann} lens models.  In doing so, we use the method described 
in \citet{Auger11} extended to multiple filters and (potentially) multiple foreground light 
distributions. To summarise, the light from neighbouring foreground galaxies is modelled 
as Sersic surface brightness distributions, as is the light from the background source.  
The same light and mass model is fit to each of the filters simultaneously, although the 
amplitudes of the surface brightness profiles are allowed to vary between the different bands. 
We also allow for small offsets between the mass and the light of the primary lensing galaxy, 
and we require the axis ratio of the mass distribution to be greater than 0.2. In all cases we 
use a single surface brightness component for the background source; when multiple sources 
are present, we model the source with the largest number of images. We are unable to find 
well-constrained models in 22 out of 51 lens systems due to the quality of the SDSS imaging; for the 
remaining 29 systems we find typical source magnifications of 5-10$\times$.  We list the 
relevant model parameters in Table 4 for those systems which we fit.    The magnification factors 
quoted refer to the ratio of total flux to intrinsic flux.   We note that our magnification 
estimates for several well-known systems (SDSS J1148+1930 and SDSS J1206+5142) are consistent (to within 
a factor of two) with previous results based off of deeper imaging (e.g., Dye et al. 2008; Jones et al. 2012b).
In order to fully exploit detailed follow-up spectroscopy, it will be crucial that we obtain higher quality imaging to 
improve the quality of the modelling in all systems.  

\subsection{Average global properties}

Equipped with the average magnification factors described in section 4.2, we can estimate the 
SFRs and stellar masses of our sample of lensed galaxies.   We will postpone a source-by-source analysis until we have acquired 
deeper optical and near-IR imaging data.   Our primary goal here is merely 
to establish the average properties likely spanned by the spectroscopic sample.  

We can crudely estimate the range of SFRs from the optical photometry from SDSS.  Given the range 
of magnification factors (typically 5-10$\times$ but as large as $\sim 30$$\times$), 
we estimate that the range of optical magnitudes of the $z\simeq 2$ subset typically translate 
into (unlensed) UV absolute magnitudes in the range  M$_{\rm{UV}} \simeq -22$ to -21.    This 
corresponds to roughly 1-3 L$^\star_{\rm{UV}}$ at $z\simeq 2$ (Reddy et al. 2012b).  To estimate 
the corresponding SFR, we need to account for dust extinction.    In absence of 
direct constraints of the rest-frame far IR continuum, this is often done through 
measurement of the reddened UV continuum slope.   Given the source selection, most of the 
systems in our sample are fairly blue with $(g-r) \simeq 0.0-0.1$ which implies UV spectral slopes in the 
range $\simeq -2.0$ to $-1.7$, somewhat bluer than average for this luminosity range (e.g., Bouwens et al. 2009; Reddy et al. 2012a).   
 Adopting the Meurer relation for our $z\simeq 2$ sample (Reddy et al. 2010; 2012a), we infer that {\it typical} star formation rates in 
the lensed SDSS $z\simeq 2$ sample likely span the range 20-100 M$_\odot$ yr$^{-1}$.   Individual systems 
within the sample will of course deviate from these values owing to the distribution functions of magnification, UV slope, 
and SFR to UV luminosity ratios present in the sample.

Knowledge of the stellar masses of the lensed galaxies will have to wait for deeper near-IR 
coverage.   While three of the sources in Table 1 are  in the Large Area Survey of the 
UKIRT Infrared Digital Sky Survey (UKIDSS), they are not detected with enough 
significance in the K-band to enable stellar mass estimates.   For the sake of comparison, we 
instead estimate the range of stellar masses of our sample through the well-established relationship between 
M$_\star$ and M$_{\rm{UV}}$ at $z\simeq 2$ (Reddy et al. 2012b).    Given our estimates of the magnification distribution  
and the observed optical magnitudes, the  stellar mass of galaxies in the $z\simeq 2$ subset is likely 
to be in the range $\simeq $10$^{10}$-10$^{11}$ M$_\odot$,  similar to the masses inferred for well-
studied examples in Table 4.    If the blue (g-r) colours in our sample reflect younger ages, this might point to 
slightly lower stellar mass to light ratios than typical objects used in the Reddy et al. (2012b) calibration, indicating 
that the lensed systems might tend to lie at the lower end of the quoted mass range.    

Clearly improved multi-wavelength imaging is a crucial requirement for fully exploiting this unique dataset.  Not only does  
such imaging provide the necessary depth to accurately characterise the observed source SED and improve estimates of total 
magnification factors, but it also delivers the necessary constraints to accurately interpret the resolved spectroscopic maps 
that we are pursuing with integral field spectrographs.  As demonstrated in our earlier work (Jones 
et al. 2012b), only with 
sufficient image quality can the source plane be reliably reconstructed.

\begin{table*}
\begin{tabular}{lcccccclll}
\hline  SDSS ID & CSWA ID &  RA & DEC &  z$_L$ & z$_S$ & R$_{\rm{ein}}$ (\arcsec) & $\mu$  &   Reference  \\ \hline

SDSS J0022+1431   &       21&    00:22:40.92 &     +14:31:10.4  &  0.380  &  2.730  &      3.3  &      9.9      &  [1]  \\
SDSS J0058$-$0721 &   102  &    00:58:48.95 &   $-$07:21:56.7  &  0.639  &  1.873  &  \ldots   &  \ldots        & This paper \\
SDSS J0105+0144   &     165   &    01:05:19.65  &    +01:44:56.4 &   0.361  &  2.127  &      3.5  &      5.4    &  This paper \\
SDSS J0143+1607   &    116 &    01:43:50.13  &    +16:07:39.0  &  0.415  &  1.499  &      2.7  &     10.7       &This paper  \\
SDSS J0145$-$0455 &  103 &    01:45:04.29  &  $-$04:55:51.6 &   0.633  &  1.958  &      1.9  &      4.7         & This paper \\
SDSS J0146$-$0929 &       22&    01:46:56.01 &  $-$09:29:52.5  &   0.440  &  1.944  &     11.9  &      9.7      & [18]  \\
SDSS J0232$-$0323 &    164 &   02:32:49.87 &   $-$03:23:26.0  &   0.450  &  2.518  &      3.7  &     20.8       & This paper \\
SDSS J0800+0812  &        11&  08:00:13.06  &     +08:12:08.4  &   0.314  &  1.408  &  \ldots   &  \ldots       &  This paper \\
SDSS J0807+4410  &   139 &   08:07:31.51  &     +44:10:48.5  &   0.449  &  2.536  &      2.1  &      3.8       & This paper \\
SDSS J0827+2232  &        23&  08:27:28.83 &      +22:32:53.9  &   0.349  &  0.766  &  \ldots   &  \ldots      & [2] \\
SDSS J0846+0446  &   141 &  08:46:47.46  &     +04:46:05.1 &    0.241  &  1.405  &      3.4  &      5.5         &This paper\\
SDSS J0851+3558  &        30 & 08:51:26.50 &      +35:58:13.8  &   0.272  &  0.919  &  \ldots   &  \ldots       & [18] \\
SDSS J0854+1008  &    142 &   08:54:28.73  &     +10:08:14.7  &   0.298  &  1.271, 1.437  &      4.2  &  4.0     &This paper \\
SDSS J0900+2234  &        19&  09:00:02.64 &      +22:34:04.9  &   0.489  &  2.033  &      7.9  &      6.5      & [3]\\
SDSS J0901+1814  &         4&  09:01:22.37 &      +18:14:32.3  &   0.346  &  2.259  &  \ldots   &  \ldots       & [4] \\
SDSS J0921+1810  &        31&  09:21:25.74  &     +18:10:17.3  &   0.683  &  1.487  &  \ldots   &  \ldots       &This paper  \\
SDSS J0952+3434  &        40&  09:52:40.22  &     +34:34:46.1  &   0.349  &  2.190  &      4.2  &      3.2      &[5] \\
SDSS J0957+0509  &        35&  09:57:39.19  &     +05:09:31.9  &   0.440  &  1.823  &      5.4  &      3.7      &[5] \\
SDSS J1002+6020  &      117 &   10:02:02.52  &   +60:20:26.3  &     0.575  &  1.114  &  \ldots  &  \ldots       &This paper  \\
SDSS J1008+1937  &        15 & 10:08:59.78  &   +19:37:17.5  &     0.306  &  2.162  &  \ldots  &  \ldots        & This paper \\
SDSS J1038+4849  &         2 & 10:38:43.58  &   +48:49:17.7  &     0.426  &  0.972, 2.20  &  \ldots  &  \ldots   & [6], [17]  \\
SDSS J1049+3544  &        33 & 10:49:23.39  &   +35:44:41.0  &     0.300  &  1.000  &      3.5  &     10.5      & [18] \\
SDSS J1110+6459  &   104 &  11:10:17.69  &   +64:59:48.2  &     0.659  &  2.481  &     11.3  &      8.1         &This paper \\
SDSS J1111+5308  &        16&  11:11:03.68  &   +53:08:54.9  &     0.412  &  1.945  &  \ldots  &  \ldots        & This paper \\
SDSS J1113+2356  &        26&  11:13:10.65  &   +23:56:39.5  &     0.336  &  0.770  &  \ldots  &  \ldots        & [7] \\
SDSS J1115+1645  &      105  &   11:15:04.39  &   +16:45:38.6 &      0.537  &  1.718  &      4.6  &      3.4    &This paper  \\
SDSS J1133+5008  &        12&  11:33:13.17  &   +50:08:40.1  &     0.394  &  1.544  &  \ldots  &  \ldots        & [8] \\
SDSS J1137+4936  &         7 & 11:37:40.06  &   +49:36:35.5  &     0.448  &  1.411  &      2.8  &      8.5      & [7] \\
SDSS J1138+2754  &        17 & 11:38:08.95  &   +27:54:30.7  &     0.447  &  0.909  &      6.2  &      4.7      & [19]  \\
SDSS J1147+3331  &   107 &   11:47:23.30  &   +33:31:53.6  &     0.212  &  1.205  &      4.6  &      8.1        &This paper \\
SDSS J1148+1930  &         1&  11:48:33.14  &   +19:30:03.1  &     0.444  &  2.379  &      5.1  &     28.7      & [9] \\
SDSS J1156+1911  &   108 &  11:56:05.46  &   +19:11:12.7  &     0.543  &  1.535  &  \ldots  &  \ldots           &This paper \\
SDSS J1206+5142  &         6 & 12:06:02.09  &   +51:42:29.5  &     0.433  &  2.000  &      3.9  &     14.9     &  [10] \\
SDSS J1207+5254  &        36&  12:07:35.91  &   +52:54:59.2  &     0.270  &  1.926  &  \ldots  &  \ldots       & [5] \\
SDSS J1209+2640  &         8 & 12:09:23.69  &   +26:40:46.7  &     0.558  &  1.018  &      8.4  &      7.0     & [11]  \\
SDSS J1226+2152  &        38 & 12:26:51.69  &   +21:52:25.5  &     0.420  &  2.923  &  \ldots  &  \ldots      & [12] \\
SDSS J1237+5533  &        13&  12:37:36.20  &   +55:33:42.9  &     0.410  &  1.864  &  \ldots  &  \ldots      & This paper  \\
SDSS J1240+4509  &         3&  12:40:32.29  &   +45:09:02.8  &     0.274  &  0.725  &      2.9  &      5.7    & [6] \\
SDSS J1244+0106  &         5&  12:44:41.01  &    +01:06:43.9  &    0.388  &  1.069  &  \ldots  &  \ldots      & [14]  \\
SDSS J1318+3942  &        37 & 13:18:11.51  &   +39:42:27.0  &     0.475  &  2.944  &  \ldots  &  \ldots      & [5]\\
SDSS J1343+4155  &        28&  13:43:32.85  &   +41:55:03.5  &     0.418  &  2.093  &  \ldots  &  \ldots      & [3] \\
SDSS J1441+1441  &        20&  14:41:49.15  &   +14:41:20.6  &     0.741  &  1.433  &      3.0  &      6.3    & [15] \\
SDSS J1450+3908  &        41&  14:50:30.65  &   +39:08:19.1  &     0.289  &  0.861  &      3.4  &      5.4    & [5] \\
SDSS J1511+4713  &        24&  15:11:18.74  &   +47:13:40.3  &     0.452  &  0.980  &      4.4  &      4.4    & [7] \\
SDSS J1527+0652  &        39&  15:27:45.02  &    +06:52:33.9  &    0.390  &  2.759  &  \ldots  &  \ldots      & [12] \\
SDSS J1629+3528  &        27 & 16:29:54.56  &   +35:28:39.5  &     0.170  &  0.850  &      3.6  &      4.1    & [7]  \\
SDSS J1723+3411  &        14&  17:23:36.16  &   +34:11:58.1  &     0.444  &  0.995  &  \ldots  &  \ldots      & [18] \\
SDSS J1958+5950  &   128 &   19:58:35.32  &   +02:57:30.2  &     0.214  &  2.225  &      6.2  &      9.3      &This paper \\
SDSS J2158+0257  &   163 &  21:58:43.68  &    +02:57:30.2  &    0.285  &  2.081  &      3.5  &      6.5       &This paper \\
SDSS J2222+2745  &   159 &  22:22:08.68  &   +27:45:35.6  &     0.485  &  2.309, 2.807  &      8.0  &   6.2    & [16], This paper  \\
SDSS J2238+1319  &        10&  22:38:31.31 &   +13:19:55.9  &      0.413  &  0.724  &  \ldots  &  \ldots      & [13] \\
SDSS J2300+2213 &111 &    23:00:17.25 & +22:13:29.7  &   0.443 &1.93 &  \ldots  & \ldots & This paper \\
\hline
\end{tabular}
\caption{Lens models of the full CASSOWARY spectroscopic sample.   Systems for which the existing data do not allow 
a reliable model are listed in the table, but no model details are provided.   
References.-- [1] Allam et al. 2007; [2] Shin et al. 2008; [3] Diehl et al. 2009; [4] Hainline et al. 2009; [5] Kubo et al. 2011; [6] 
Belokurov et al. 2009; [7] Kubo et al. 2009; [8] Sand et al. 2005; [9] Belokurov et al. 2007; [10] Lin et al. 2008; [11] Ofek et al. 2008; 
[12] Koester et al. 2010; [13] Bayliss et al. 2011; [14] Christensen et al. 2010; [15] Pettini et al. 2009; [16] Dahle et al. 2012; 
[17] Jones et al. (2013), [18] CASSOWARY data release; [19] Wuyts et al. 2012.
}
\end{table*}

\section{Discussion}

The sample presented in this paper will provide a major stimulus to efforts to use 
lensed galaxies as detailed probes of galaxy formation at high redshift.    
Detailed optical and near-infrared follow-up of this new population is now underway 
and will be discussed in future papers.  In the following we provide an initial exploitation of the sample 
using the existing data.

\begin{figure*}
\begin{center}
\includegraphics[width=0.49 \textwidth]{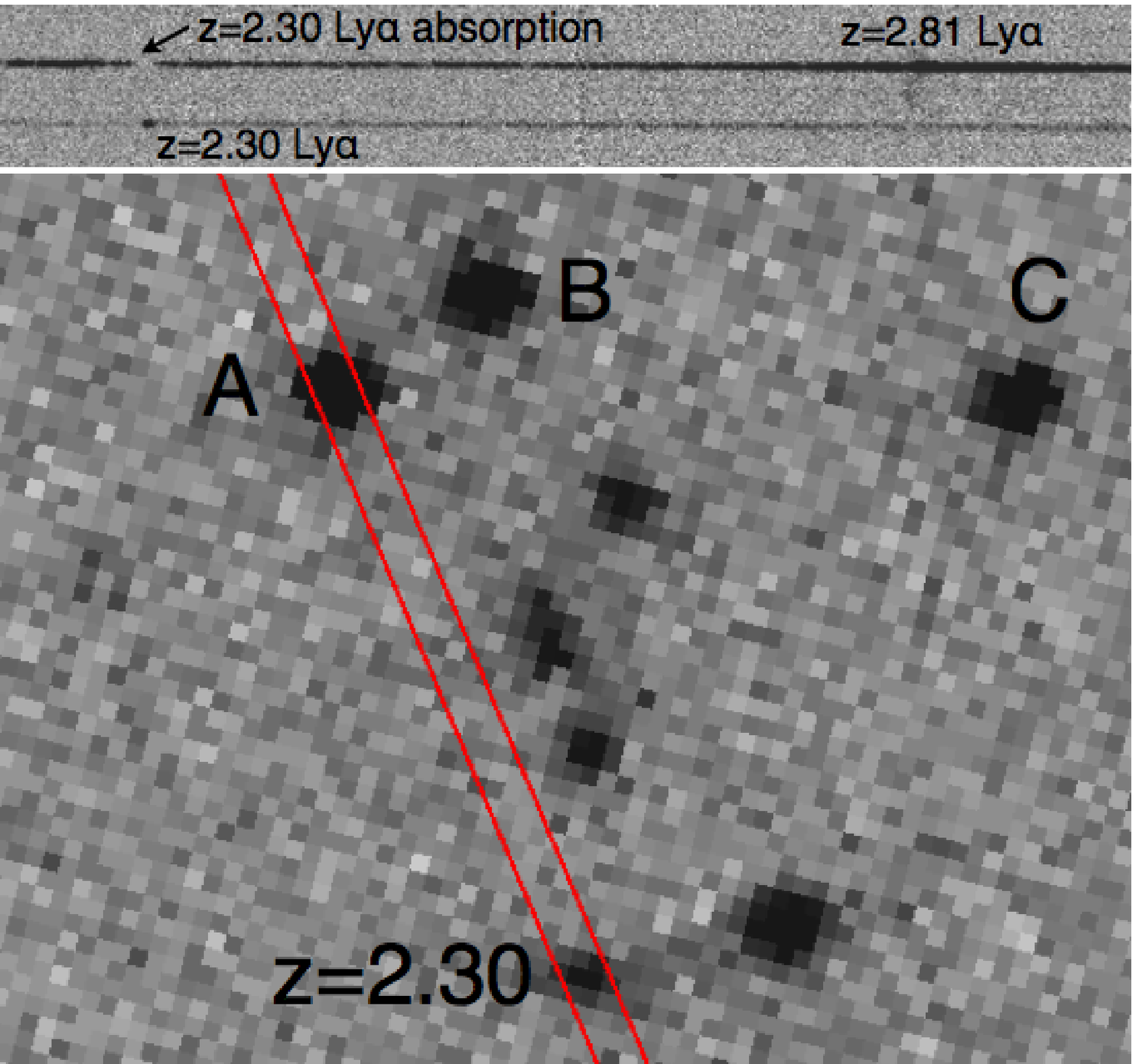}
\includegraphics[width=0.47 \textwidth]{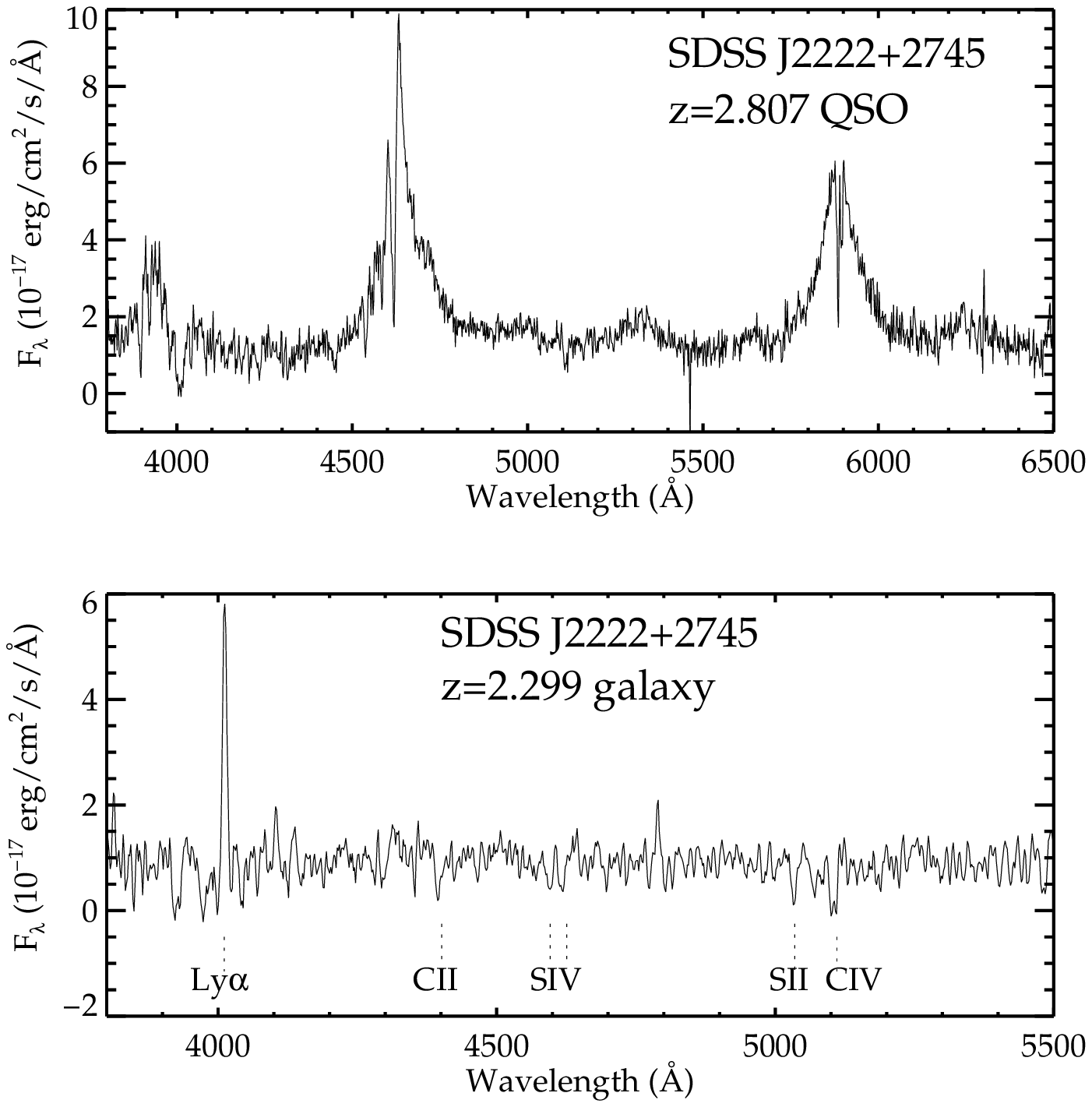}
\caption{The bright lensed $z=2.807$ quasar images in SDSSJ2222+2745 enable the 
opportunity to characterise the kinematics and metal content of the gas within $\simeq 50$ kpc of 
a bright lensed $z=2.299$ galaxy.    ({\it Left:})  
The SDSS g-band image (bottom panel) clearly shows three of the six known images of the 
$z=2.807$ quasar (which we labeled as A, B, and C) and a $z=2.299$ lensed galaxy.   We have 
obtained three separate MMT spectra of this system, one of which is denoted by the red lines.   
A portion of one of the 2D MMT spectrum is shown at top.  Clearly visible is Ly$\alpha$ emission 
from the galaxy (bottom spectrum) and extended Ly$\alpha$ emission from the quasar (top).   
Notice the complete absorption of the quasar continuum at the observed wavelength 
corresponding to Ly$\alpha$ in the frame of the $z=2.299$ galaxy.
  ({\it Right:}) Extracted MMT spectra of quasar image A (top) and lensed galaxy (bottom).   
  Absorption of the quasar continuum near 4010~\AA\ indicates the presence of neutral hydrogen  
  at an impact parameter of roughly $\simeq 50$ kpc from the $z=2.299$ galaxy.  
    }
\label{fig:absspectra}
\end{center}
\end{figure*}

\begin{figure}
\begin{center}
\includegraphics[width=0.47 \textwidth]{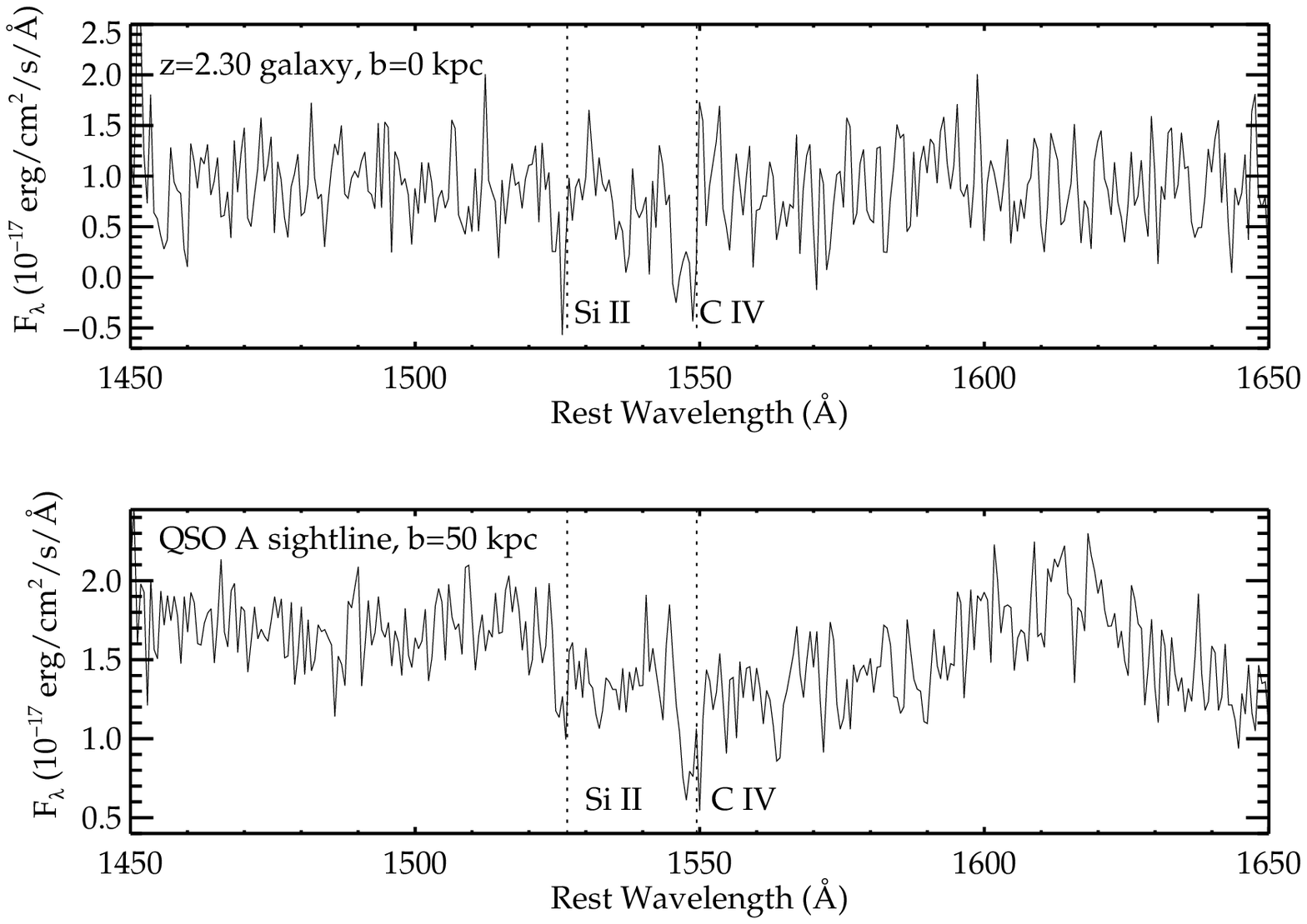}
\caption{Spectra illustrating spatial extent of Si II and CIV gas associated with lensed $z=2.299$ galaxy 
in SDSS J2222+2745.    ({\it Top:}) Si II $\lambda$1526 and CIV$\lambda$1549 absorption 
along the line of sight toward the $z=2.299$ galaxy.   ({\it Bottom:})  Spectrum of the $z=2.807$ quasar shifted to the rest-frame of the $z=2.299$ galaxy.   The quasar continuum shows 
absorption from Si II $\lambda$1526 and CIV $\lambda$1549, highlighting the presence of metals at 
an impact of $\sim 50$ kpc.}
\label{fig:absspectra}
\end{center}
\end{figure}

\subsection{A lensed extreme emission line galaxy}

Most of the optically-bright lensed $z\simeq 2-3$ galaxies that have been studied in 
detail  are broadly similar in their properties to typical systems within the population of 
UV-continuum selected galaxies (e.g., Shapley 2011 and references therein).  But as the 
sample of bright lensed galaxies increases, we expect to find outliers that offer unique 
insight into galaxy formation at high redshift.   

The $z=1.43$ arc in SDSS J0846+9446 appears to be such a source.   While this is 
by no means the brightest continuum source in our sample, the EWs of the Balmer 
lines are much larger than are typically seen in UV-selected galaxies (see Figure 3), 
allowing us to detect H$\alpha$ 
through H12 with the short 20 min integration detailed in Table 1.   The flux ratio of H$\alpha$ and 
H$\beta$ (2.79) suggest that the nebular gas is largely unattenuated by dust, while the rest-frame EW of H$\alpha$ emission 
($\gsim 300$ \AA) points to ionising radiation from a reasonably young ($\lsim 100$ Myr for constant star formation history 
using nebular emission models described in Robertson et al. 2010) 
stellar population.   The detection of [OIII] $\lambda$4363 (Figure 3) enables a measurement of the electron 
temperature (T$_e \simeq 1.5\times$10$^{4}$ K) which can be used along with the measured line ratios of the oxygen lines 
(following Izotov et al. 2006) to compute the oxygen abundance of the gas.   
Not surprisingly given the young age and lack of dust, SDSS J0846+9446 
appears to be one of the more metal poor galaxies known at high redshift with 12+log(O/H)$ 
\simeq 8.0 $ ($\sim 0.2$ Z$_\odot$).  

The lower limit we place on the [OIII] $\lambda$5007 rest-frame EW ($>$320~\AA) is in the range 
spanned by the population of extreme line emitting galaxies (EELGs) identified in imaging and 
grism surveys (e.g., Kakazu et al. 2007; Hu et al. 2009; Atek et al. 2010 ; van der Wel et al. 2012).  
The number density of EELGs at $z\simeq 1.5-2$ is large enough to indicate that they are 
likely an important mode of star formation among low mass galaxies at high redshift and may well 
represent the manner in which the stellar mass in dwarf galaxies is assembled (e.g., van der Wel et. 2012).   
Yet as a result of their faint continuum flux densities, our understanding of the detailed 
properties of extreme line emitters at $z>1$ remains limited (see Hu et al. 2009 for a detailed discussion 
at $z\simeq 0-1$).

By finding a highly-magnified example of such a system, we can begin to study this population 
in far more detail than would be possible in unlensed examples at $z\simeq 1.5-2.0$.   With the 
spectra we have acquired to date for SDSS J0846+9446, we will be able to explore the relative 
chemical abundances in the ionised gas and to characterise the properties of the outflowing 
gas and stellar populations.   Such an in depth analysis is beyond the scope of the present catalog paper 
and will appear in a separate work.   

\subsection{Spatial extent of gas around a $z=2.3$ galaxy}

The rest-UV absorption line spectra presented in Figure 2 provide information on the 
kinematics and chemical content of outflowing gas in galaxies at high redshift but do not reveal 
the exact location of the gas along the line of sight.   Without 
knowledge of the spatial extent of the outflowing material, it is virtually impossible to characterise the 
energetics associated with the outflows.   Spectra of bright background galaxies or quasars near 
the line of sight of a given galaxy provide a means of overcoming this shortcoming.    

Steidel et al. (2010) conducted a comprehensive analysis of composite spectra of galaxy pairs, providing 
constraints on the distribution and covering fraction of circumgalactic gas on physical scales 
spanning 3-125 kpc.     The stacked spectra revealed metal line absorption at distances of $\simeq 60$ kpc and 
Ly$\alpha$ absorption from neutral hydrogen on physical scales in excess of 100 kpc.   

A logical goal for the future is to extend constants on the spatial extent of outflows to individual systems.   
Most unlensed galaxy pairs are much too faint for useful constraints on the spatial extent of outflows.  
Steidel et al. (2010) present a case study for one of the brightest pairs in their sample, revealing CIV and Al II 
absorbing gas at an impact parameter of 16 kpc from a $z=1.61$ galaxy.   The continuum 
magnitudes of this particularly pair ($g'$=23.5 and 23.7) are $\simeq 1$ magnitude brighter than typical galaxies in 
their sample, making this a unique measurement at $z\simeq 2$.

Lensing clusters offer a feasible way of constraining the spatial distribution of circumgalactic gas on a source 
by source basis.    The cluster environment offers two advantages.   First the surface density of background 
sources is enhanced, providing more probes of the spatial distribution of gas associated with any 
individual lensed galaxy.   Second, the flux magnification provided to lensed systems enables much higher S/N 
(and higher resolution) spectra to be obtained.    With constraints from multiple impact parameters per galaxy 
(at various position angles), it should be possible to characterise the spatial and angular distribution of circumgalactic
gas associated with an individual lensed galaxy.    But with these advantages also comes 
difficulty, as knowledge of the impact parameters of the lensed background galaxies can only be obtained with a reliable cluster mass model, limiting analysis to well-constrained lenses.  

The close proximity of a bright ($g\simeq 21$ as measured in deeper imaging presented in Dahle et al. 2012) lensed $z=2.30$ galaxy to three bright images of a lensed $z=2.81$ 
quasar in SDSS J2222+2745 (Figure 4) enables immediate constraints on the circumgalactic gas associated with 
the foreground $z=2.30$ galaxy.   The brightest quasar images are separated by 14, 15, and 16 arc seconds with 
respect to the galaxy.  But the relevant quantity for calculating the impact parameter is of course the separation 
in the source plane of the $z=2.30$ galaxy.    Using our lens model for this cluster (see \S4.2), we find that 
the quasar absorption line spectra probes gas associated with the $z=2.30$ galaxy on physical 
scales of 50 kpc.   While each quasar image probes the same impact parameter, the individual spectra can 
be stacked allowing fainter absorption features to be detected.    

The relatively shallow spectra we have obtained with MMT provide a first glimpse of the spatial extent 
of circumgalactic gas around the lensed $z=2.30$ galaxy.   It is immediately apparent in the 2D spectrum that the quasar 
continuum in image A is completely extinguished at the wavelength corresponding to Ly$\alpha$ in the 
rest-frame of the $z=2.30$ galaxy.   
The centroid of the $z=2.30$ Ly$\alpha$ absorption in the quasar spectrum 
occurs at roughly 4010~\AA\ with strong absorption spanning $\sim 4000$-4020~\AA.  This is not surprisingly somewhat bluer than the wavelength of the Ly$\alpha$ 
emission(4012 ~\AA)  in the galaxy spectrum.    Metal absorption from the CGM of the $z=2.30$ galaxy is 
also apparent in quasar spectrum.   We clearly detect CIV absorption at $\sim 5110$~\AA\ and tentatively identify 
absorption from Si II $\lambda$1526 at 5034~\AA\ in the quasar spectrum (Figure 5).  Higher S/N and higher 
resolution observations are required to characterise the profiles of the absorption lines in both the quasar and galaxy.

Deeper spectra of the quasar images should reveal the presence of other ions associated with the $z=2.3$ galaxy 
at the 50 kpc impact parameter, providing a more detailed picture of the chemical makeup and 
ionisation state of material associated with the galaxy.   Meanwhile as deeper multi-wavelength 
images of the field around SDSSJ2222+2745 emerge, it should be possible to identify additional lensed 
systems behind the $z=2.30$ galaxy.    Spectra of these systems should yield a map of the spatial 
and angular distribution of outflowing gas associated with this galaxy.

\subsection{Intervening absorption from foreground lens}

Deep spectra of the background lensed galaxies can also provide information on the CGM of the quiescent central 
lens galaxy.   In particular, 
if a significant component of cool gas resides in the halos of the early-type lens galaxies (as expected from the 
QSO absorption line results presented in Werk et al. 2012), then we should 
detect absorption from Mg II or Fe II transitions in the spectra of the background source.   Given the range of angular 
separations between arcs and lenses, the background lensed galaxies probe impact parameters of $\simeq 20-60$ kpc.

To detect metal line absorption in the background lensed source requires that the lens be at high enough 
redshift to place either Mg II or Fe II transitions in the optical window.   This limits the test to lenses at 
$z\gsim 0.35$ (for Mg II) and $z\gsim 0.45$ (for Fe II $\lambda$2600). As is clear from Table 4, many of the 
CASSOWARY systems satisfy these criteria.  

We have examined the MMT source spectra  for possible signs of absorption from cool gas associated with the 
lens.   As noted in \S3.2, there are two cases in which we suggest that intervening absorption might be associated 
with the lens.   In SDSS J0145-0455 (Figure 2), the absorption feature at 4244~\AA\ might correspond to 
Fe II $\lambda$2600 absorption from the central lens.    We should be able to confirm this identification through 
detection of Mg II absorption, but given the limited resolution of the MMT spectra, Mg II absorption at the lens redshift 
is blended with the broad CIV absorption from the background arc.   A higher resolution spectrum is needed to 
confirm the nature of the absorption system.   In the arc toward SDSS J1439+3250, we detect absorption 
at 3970~\AA, which may stem from Mg II at the redshift ($z=0.418$) of the lens.  This identification is complicated by the fact 
that we have not yet confirmed the redshift of the arc, making it impossible at this stage to disentangle whether 
a given absorption feature stems from the arc or lens.  Several other systems with MMT spectra show hints of absorption at the wavelength 
expected for Mg II, but as a result of our limited S/N and low spectral resolution, it is not yet possible to make definitive 
identifications.   

As higher S/N and higher resolution spectra of the lensed galaxies emerge, it should become feasible to put 
constraints on the properties of cool gas associated with quiescent galaxies at $z\simeq 0.4-0.7$.  In the ideal 
scenario in which the background source forms a complete Einstein ring, such absorption line studies 
offer the potential to create a full two-dimensional map (at a single impact parameter) of the CGM of the lens, 
complementing ongoing work with quasar absorption lines (e.g., Chen et al. 2010; Werk et al. 2012).  

\section{Summary}

The discovery of bright gravitationally-lensed galaxies has begun to make a significant impact on our understanding 
of star formation and feedback at high redshift.    Once a field focused on the 
few bright highly-magnified galaxies known (e.g., Pettini et al. 2001), sky  
surveys like SDSS and targeted HST imaging of massive clusters have opened the door for 
the construction of much larger samples of bright lensed galaxies.   For several years, our team 
(CASSOWARY) has been identifying  galaxy-galaxy lens candidates within 
the SDSS imaging data (e.g., Belokurov et al. 2007, 2009).   Yet the impact of detailed follow-up efforts has
been stunted by the low percentage of the brightest lensed systems  with confirmed redshifts.   

In this paper, we present the results of a spectroscopic campaign aimed at confirming the 
redshifts of a larger fraction of the CASSOWARY lensed candidates identified in SDSS.  Our 
ultimate goal is to increase the number of suitable targets for resolved IFU infrared spectroscopy 
(e.g., Stark et al. 2008; Jones et al. 2012b) and high resolution optical spectroscopy (Quider et al. 
2009; 2010).  

The spectroscopic data used to confirm redshifts in this paper were primarily obtained using the Blue and Red 
Channel Spectrographs on MMT, but we also 
observed select objects with spectrographs on Keck, LBT, and Magellan.     Through these 
efforts, we have confirmed the redshifts 
of 25 bright lensed objects (24 galaxies and 1 quasar) in SDSS.   Only two of the lensed systems (the galaxy 
and quasar behind SDSS 2222+2745 and the $z=0.909$ galaxy toward SDSS J1138+2754) have been 
presented with redshifts (Dahle et al. 2012; Wuyts et al. 2012).   The new redshift sample contains two  
of the brightest ($r\lsim 20$) high redshift galaxies known in SDSS, a young, metal poor (0.2 Z$_\odot$)  
extreme line emitting star forming system at $z=1.43$, and many further systems which are bright enough for 
detailed studies.   We provide a brief discussion of the 
spectroscopic properties of each system, focusing on the absorption or emission features used for 
redshift confirmation.   

With the new redshift catalog presented in this paper, the total number of spectroscopically-confirmed lensed 
systems in the CASSOWARY SDSS database now stands at 55.  Our initial attempts at lens modelling of the CASSOWARY  
systems suggests that source magnification factors are typically in the range of 5-10$\times$.   Given the 
range of SDSS apparent magnitudes of the sources, we crudely estimate typical star formation rates (20-100 M$_\odot$ yr$^{-1}$ ) and stellar masses (10$^{10}$ - 10$^{11}$ M$_\odot$) of the $z\simeq 2$ lensed population.  Deeper imaging will be required to extend these estimates robustly to individual systems.  Detailed spectroscopic follow-up of these systems is underway and will be presented in future papers.  
With 26 bright lensed systems in the CASSOWARY catalog  at $z\gsim 1.8$ and a further 10  at $1.3\lsim z\lsim 1.8$, 
it will soon be possible to infer more generalised conclusions from detailed spectroscopic study of the brightest lensed 
galaxies.  

\section*{Acknowledgments}
We thank Fuyan Bian, Zheng Cai, Linhua Jiang, Ian McGreer, and Evan Schneider for conducting some of the 
observations used in this paper.  DPS acknowledges support from NASA through Hubble Fellowship 
grant \#HST-HF-51299.01 awarded by the Space Telescope Science Institute, which is operated by the 
Association of Universities for Research in Astronomy, Inc, for NASA under contract NAS5-265555.   BER 
is supported by Steward Observatory and the University of Arizona College of Science.
Observations reported here were obtained at the MMT Observatory, a joint facility of the University of Arizona and the Smithsonian Institution. 

\footnotesize{
  \bibliographystyle{mn2e}
  \bibliography{paper}
}

\label{lastpage}

\end{document}